\documentclass[%
 reprint,
 amsmath,amssymb,
 aps,
]{revtex4-2}

\usepackage{lineno}
\usepackage{graphicx}
\usepackage{dcolumn}
\usepackage{bm}

\usepackage{color}
\usepackage{xspace}
\usepackage[
colorlinks=true,
filecolor=black,
anchorcolor=blue,
linkcolor=blue,
citecolor=cyan, 
urlcolor=cyan,
linktocpage=true,
plainpages=false,
breaklinks=true,
pdfstartview=FitH
]{hyperref}
\usepackage[capitalise]{cleveref}
\usepackage{textgreek}

\def\meth{\texttt{\textSigma Track}\xspace}

\begin{document}


\title{Methods for Detecting Gravitational Waves from mini-Extreme-Mass-Ratio Inspirals \uppercase\expandafter{\romannumeral 2}: A Spectral-Leakage-Aware Framework }

\author{Zi-Xuan Wang$^{1,2}$}
\email{wangzixuan243@mails.ucas.ac.cn}
\author{Xing-Yu Chen$^{3}$}%
\email{chenxingyu223@mails.ucas.ac.cn}
\author{Ju Chen$^{1,2}$}
\email{chenju@ucas.ac.cn}
\author{Gong Cheng$^{1,2}$}
\email{chenggong@ucas.ac.cn}
\author{Huai-Ke Guo$^{1,2}$}
\email{guohuaike@ucas.ac.cn}
\author{Andrew L. Miller$^{1,2,4,5}$}
\email{andrewlawrence.miller@ligo.org}
\affiliation{$^{1}$International Center for Theoretical Physics Asia-Pacific (ICTP-AP), University of Chinese Academy of Sciences, Beijing 100190, China}
\affiliation{$^{2}$Taiji Laboratory for Gravitational Wave Universe, University of Chinese Academy of Sciences, 100049 Beijing, China}
\affiliation{$^{3}$Department of Physics, University of Chinese Academy of Sciences, Beijing 100190, China}
\affiliation{$^{4}$Nikhef -- National Institute for Subatomic Physics,
Science Park 105, 1098 XG Amsterdam, The Netherlands}
\affiliation{$^{5}$Institute for Gravitational and Subatomic Physics (GRASP),
Utrecht University, Princetonplein 1, 3584 CC Utrecht, The Netherlands}




\date{\today}


\begin{abstract}
Mini-Extreme-Mass-Ratio Inspirals (mini-EMRIs), comprising a sub-solar exotic compact object (such as a primordial black hole or boson star) orbiting a much heavier stellar-origin or exotic compact object, represent key targets for ground-based gravitational-wave detectors to probe the early universe and the nature of dark matter. 
However, detecting such systems, which could spend hours to years in LIGO, Virgo and KAGRA data, poses a computational challenge to standard matched-filtering methods, 
which has motivated the development of methods that divide the data into coherent chunks, Fourier transform them, and combine them incoherently. However, these 
pipelines typically require that the signal be quasi-monochromatic within each chunk, 
which imposes a fundamental limit on the coherence time, since spectral leakage becomes significant when the frequency evolution within the chunk becomes comparable to the frequency resolution. In this work, we extend the development of our method, \texttt{\textSigma Track} \cite{MyUnpublishedPaper}, to the regime in which the quasi-monochromatic approximation is relaxed,
in two ways. First, we establish an analytical model for the spectral leakage, extending the validity of conventional analyses beyond the quasi-monochromatic regime. Second, we propose the $\Sigma R$ statistic---a novel detection metric formed by a weighted summation of power ratios---which effectively recovers the signal energy dispersed across adjacent frequency bins. Building on this framework, we further introduce an innovative frequency-layered search strategy that dynamically optimizes the coherence time across the observation band. 
We benchmark our method against a globally optimized Hough transform pipeline using a fiducial mini-EMRI signal from a binary with masses $(1.5,10^{-5})\,M_\odot$. The results demonstrate that our framework achieves an order-of-magnitude enhancement in the effective detection volume, significantly expanding the horizon for discovering mini-EMRIs and sub-solar exotic compact objects with ground-based gravitational wave detectors, which can be similarly applied to EMRI searches for future space-based gravitational wave detectors.
\end{abstract}

\maketitle


\section{Introduction}
\label{sec:intro}


The advent of gravitational-wave (GW) astronomy, heralded by the first direct detection of a binary black hole merger \cite{LIGOScientific:2016aoc} and enabled by the global network of interferometric detectors---Advanced LIGO \cite{LIGOScientific:2014pky}, Advanced Virgo \cite{VIRGO:2014yos}, and KAGRA \cite{KAGRA:2018plz}---has provided a pristine probe into the universe's most energetic phenomena. The rapidly expanding catalog of transient events now encompasses hundreds of compact binary coalescences (CBCs) \cite{KAGRA:2021vkt,LIGOScientific:2025slb,Nitz:2021uxj,Nitz:2021zwj}. While these observations have profoundly advanced our understanding of compact object formation, they have simultaneously exposed significant tensions with established astrophysical models. 
The observed population exhibits properties challenging to reconcile with canonical stellar evolution theories \cite{Belczynski_2010}, ranging from detections within the presumed high-mass gap \cite{LIGOScientific:2020iuh} to the discovery of enigmatic sources populating the ``lower mass gap'', whose existence obscures the conventional distinction between neutron stars and black holes \cite{LIGOScientific:2020aai,LIGOScientific:2020zkf,LIGOScientific:2021qlt,Farr_2011}.
These observational puzzles imply the existence of alternative formation pathways, beyond astrophysical processes \cite{PortegiesZwart:1999nm, Rodriguez:2015oxa,Gerosa:2021mno,Mandel:2021smh}. Moreover, a smoking gun evidence for new physics is the detection of a sub-solar compact object, which has received increasingly rising
attention in the community, resulting in many searches for such objects \cite{LIGOScientific:2018glc,LIGOScientific:2019kan,Singh:2020wiq,Nitz:2020bdb,LIGOScientific:2021job,Phukon:2021cus,Nitz:2021mzz,Nitz:2021vqh,LVK:2022ydq,LIGOScientific:2022hai,Nitz:2022ltl,Miller:2024fpo,LIGOScientific:2025vwc}. Popular candidates for the sub-solar exotic compact
objects are primordial black holes \cite{Hawking:1971ei,Khlopov:2008qy,Carr:2016drx,Carr:2019kxo, Escriva:2022duf}, boson stars~\cite{Liebling:2012fv}, etc, the detection of which can also shed new
light in the understanding of the nature of dark matter.

For binaries consisting of increasingly light sub-solar exotic compact objects, the gravitational
wave signal decreases significantly as the chirp mass becomes smaller. An optimal strategy is to pair a 
sub-solar exotic compact object up with a much heavier one (of stellar or compact origin) to increase the signal strength, leading to a sytem with an extreme mass ratio $q = m_\mathrm{lighter} / m_\mathrm{primary} \lesssim 10^{-5}$, thus called mini extreme mass ratio inspirals (mini-EMRIs)  \cite{Guo:2022sdd}. While comparable-mass binaries ($q \sim 1$) are now routine targets for ground-based detectors \cite{maggiore_gravitational_2008,Usman:2015kfa} and classical EMRIs involving supermassive black holes are key objectives for future space-based missions such as LISA \cite{LISA:2017pwj,Berry:2019wgg,Gair:2008bx,Gair:2017ynp}, Taiji \cite{Hu:2017mde,Ruan:2018tsw,Wu:2018clg} and TianQin \cite{TianQin:2015yph,TianQin:2020hid,Luo:2020bls}, mini-EMRIs serve as a critical phenomenological bridge. They offer the unprecedented opportunity to study extreme mass-ratio dynamics within the sensitive frequency band of terrestrial interferometers, ultimately opening a novel window into fundamental physics.

The signal phenomenology of a mini-EMRI diverges fundamentally from that of canonical compact binary coalescences (CBCs). In contrast to the fleeting ``chirps'' characteristic of stellar-mass mergers, a mini-EMRI trajectory may linger within the detector's most sensitive frequency band for timescales spanning days to years. This ``long-transient'' character places mini-EMRIs in a hybrid signal category: they exhibit the deterministic frequency evolution typical of inspiraling binaries, yet possess the temporal persistence of continuous wave (CW) sources. This duality imposes severe constraints on standard data analysis strategies. On one hand, the computational cost of fully coherent matched filtering becomes prohibitive due to the rapid expansion of the template bank parameter space over such extended durations \cite{Owen:1995tm,Brady:1997ji,Walsh:2016hyc,Keitel:2015ova,PhysRevLett.127.151101}. On the other hand, conventional CW search pipelines---optimized for quasi-monochromatic sources like spinning neutron stars---lack the dynamic range to accommodate the significant frequency drift inherent to mini-EMRI dynamics \cite{PhysRevD.70.082001,Astone_2014,Piccinni:2018akm,Astone_2005,Miller:2018rbg,Miller:2024jpo,Menon:2025wce,Miller:2025ote}. Consequently, the detection of these sources necessitates the development of specialized search algorithms designed to efficiently track rapid phase evolution over long observational baselines.


Analyses of such non-stationary signals typically rely on the Short-Time Fourier Transform (STFT) as a fundamental preprocessing step to track time-frequency evolution. However, a fundamental limitation arises when the signal frequency evolves significantly within a single analysis window, inducing spectral leakage that degrades detection efficiency. Standard STFT-based algorithms, such as the Hough transform \cite{PhysRevD.70.082001,Astone_2014}, operate under a ``quasi-monochromatic'' approximation, assuming the signal remains effectively monochromatic within a segment of duration $T$. This constraint is succinctly quantified by the dimensionless widening factor $w = |\dot{f}|T^2$, which is empirically required to be $w \lesssim 0.5$---implying that the signal's frequency drift over the segment duration $T$ is confined to less than half of the frequency resolution $f_\mathrm{bin} = 1/T$. For a long-transient GW signal, this condition constrains the choice of the coherent time. While a longer $T$ is desirable for enhancing the signal-to-noise ratio (SNR) through extended coherent integration, analysts are compelled to use a short $T$ to satisfy the approximation. This creates a fundamental trade-off: maximizing detection sensitivity is sacrificed for model fidelity, thus curtailing the reach for more distant or fainter sources.


This work establishes a theoretical framework to rigorously model spectral leakage, thereby enabling the unconstrained selection of the coherence time $T$, independent of the traditional quasi-monochromatic restriction. Leveraged by the massive parallelism of modern Graphics Processing Unit (GPU) architectures \cite{Tenorio:2024jgc,Merou:2025ark}, which allows computationally intensive approaches to rival the efficiency of the Hough transform, we build upon the foundations of the PowerFlux method \cite{LIGOScientific:2007hnj} to introduce a novel detection statistic, denoted as $\Sigma R$. We fully characterize the statistical properties of $\Sigma R$, providing a theoretical guide for the optimization of search parameters that effectively resolves the trade-off between coherence time and spectral leakage loss. Ultimately, our method achieves an order-of-magnitude improvement in detection volume compared to the conventional Hough transform pipeline.


The structure of this paper is as follows. \cref{sec:model} delineates the mini-EMRI signal morphology and the corresponding interferometric detector response. Confronting the limitations of standard time-frequency analysis, \cref{sec:Analysis} develops a rigorous analytical framework within the STFT domain to precisely model the phenomenon of spectral leakage. Building on this foundation, \cref{sec:statistic} formally introduces the $\Sigma R$ statistic, deriving its statistical properties and establishing analytical metrics for both the maximum detectable distance and the effective parameter space volume. Finally, in \cref{sec:result}, we validate the performance of our algorithm through comprehensive simulations and present a comparative benchmark against established pipelines. 

\section{Signal model}
\label{sec:model}
As described in \cite{Jaranowski:1998qm,Riles:2022wwz}, the response $s(t)$ of an interferometric gravitational-wave detector to an incoming signal is a linear combination of the two gravitational-wave polarizations, $h_{+}(t)$ and $h_{\times}(t)$:
\begin{equation}
s(t) = F_{+}(t)\,h_{+}(t) + F_{\times}(t)\,h_{\times}(t),
\label{eq:signal_response}
\end{equation}
where $F_{+}(t)$ and $F_{\times}(t)$ are the detector antenna-pattern functions (they depend on the detector orientation, source orientation and the source sky location).

The polarization waveforms for a quasi-sinusoidal signal can be written as
\begin{equation}
\begin{aligned}
h_{+}(t) &= h_0(t)\,\frac{1+\cos^2\iota}{2}\,\cos\Phi(t),\\
h_{\times}(t) &= h_0(t)\,\cos\iota\,\sin\Phi(t),
\end{aligned}
\end{equation}
where $h_0(t)$ denotes the intrinsic strain amplitude, $\Phi(t)$ the phase, and $\iota$ the inclination angle between the orbital angular momentum and the line of sight. The intrinsic amplitude $h_0(t)$ is in general a slowly varying function of time through its dependence on frequency $f(t)$.

For clarity of the subsequent analysis, it is convenient to combine the two polarization contributions into a single sinusoid:
\begin{equation}
s(t) = h_0(t)\,Q(t)\,\cos\big[\Phi(t)+\phi_p(t)\big],
\label{eq:signal_compact}
\end{equation}
where the dimensionless modulation factor $Q(t)\ge0$ collects the time-dependent projection of the polarizations onto the detector:
\begin{equation}
Q(t) \equiv \sqrt{ F_{+}^2(t)\left(\frac{1+\cos^2\iota}{2}\right)^{\!2} + F_{\times}^2(t)\cos^2\iota }.
\end{equation}
The detector-frame mixing of polarizations also produces a (slowly varying) polarization phase
\begin{equation}
\phi_p(t) \equiv -\arctan\!\left(\frac{2\cos\iota\,F_{\times}(t)}{(1+\cos^2\iota)\,F_{+}(t)}\right),
\end{equation}
chosen so that \cref{eq:signal_compact} reproduces \cref{eq:signal_response}.

In this work we model a mini-EMRI as a compact object in a circular, equatorial orbit about a (possibly spinning) central body. Under this assumption the spin-up $\dot f$ and the instantaneous intrinsic amplitude $h_0$ can be written as power-law forms modified by relativistic correction factors \cite{Guo:2022sdd,Finn:2000sy}:
\begin{equation}
\begin{aligned}
&\dot{f} = \frac{96}{5}\,\pi^{8/3}\!\left(\frac{G\mathcal{M}_c}{c^3}\right)^{\!5/3} f^{11/3}\;C_f(a,f),\\[4pt]
&h_0(f) = \frac{4}{d}\!\left(\frac{G\mathcal{M}_c}{c^2}\right)^{\!5/3}\!\left(\frac{\pi f}{c}\right)^{\!2/3}\;C_h(a,f),
\label{eq:relativistic}
\end{aligned}
\end{equation}
where $\mathcal{M}_c=(m_1m_2)^{3/5}(m_1+m_2)^{-1/5}$ is the chirp mass, $d$ the luminosity distance, and $a$ the dimensionless spin of the central body. 

The correction factors $C_f(a,f)$ and $C_h(a,f)$ encode relativistic (strong-field) effects introduced by the central body's spin and by higher-order post-Newtonian physics. In the weak-field, large-separation limits $C_f\to1$ and $C_h\to1$, so \cref{eq:relativistic} reduce to the standard Newtonian power-law expressions. As the orbit approaches the innermost stable circular orbit (ISCO), these correction factors depart significantly from unity and capture the rapid relativistic evolution of both phase and amplitude.

Finally, to obtain the phase as measured in the detector's reference frame, we must account for the modulation induced by the detector's motion. The dominant effect is the Doppler shift, which modifies the signal's instantaneous frequency. Integrating this observed frequency yields the detector-frame phase $\Phi(t)$:
\begin{equation}
\Phi(t) = 2\pi\int^{t} f(t')\left(1+\frac{\vec{v}(t') \cdot \hat{n}}{c}\right)\,dt' + \Phi_0,
\label{eq:phase_doppler}
\end{equation}
with $\Phi_0$ a reference phase and $\vec{v}(t') \cdot \hat{n}$ represents the component of the detector's velocity along the line of sight to the source. The combination of this phase model (\cref{eq:phase_doppler}) with the amplitude evolution described by \cref{eq:signal_compact,eq:relativistic} provides the complete time-domain signal model used to construct the time-frequency templates in our semi-coherent search.

\section{Spectral Analysis}
\label{sec:Analysis}

\subsection{Short-time Fourier Transform}

The strain data recorded by the interferometer is denoted as $x(t)$ and is modeled as the linear superposition of a deterministic gravitational-wave signal $s(t)$ and stochastic instrumental noise $n(t)$:
\begin{equation}
    x(t) = s(t) + n(t).
    \label{eq:data_model}
\end{equation}
Throughout this article, we approximate the noise component $n(t)$ as a stationary, Gaussian random process.

To analyze the time-frequency evolution of the signal, the discrete data stream is partitioned into overlapping segments of duration $T_\mathrm{DFT}$, each comprising $N$ samples. A window function, $W[n]$, is applied to each segment prior to the computation of the Discrete Fourier Transform (DFT) to mitigate spectral leakage at the boundaries. We enforce a power-normalization on the window function, a detailed derivation and discussion of this convention are provided in \cref{app:window_function_power}. For the $i$-th time segment, the complex-valued STFT coefficient at the $k$-th frequency bin is given by:
\begin{equation}
    \tilde{x}_i[k] = \frac{1}{N}\sum_{n=0}^{N-1} x_i[n] W[n] e^{-i2\pi nk/N},
\end{equation}
where $x_i[n]$ denotes the discrete time samples within the $i$-th segment. This procedure is standard in CW analyses, e.g. the construction of short FFTs and short fast Fourier transform databases \cite{Astone_2005}.

To facilitate statistical detection, we define a normalized \textit{power-ratio spectrogram}, $R[i,k]$, by rescaling the measured power in each time-frequency bin by the expected noise power:
\begin{equation}
    R[i,k] \equiv \frac{|\tilde{x}_i[k]|^2}{\langle|\tilde{n}_i[k]|^2\rangle}.
    \label{eq:ratio}
\end{equation}
Here, the denominator $\langle|\tilde{n}_i[k]|^2\rangle$ represents the estimated value of the power spectrum of the noise background in the $(i,k)$-th bin. In practice, this baseline is estimated from the data itself using robust averaging techniques, such as the auto-regressive average spectrum method, to exclude transient outliers \cite{Kay1988ModernSE}.



Under the assumption of stationary Gaussian noise, the scaled statistic $2R[i,k]$ follows a non-central chi-squared distribution with two degrees of freedom:
\begin{equation}
    2R[i,k] \sim \chi^2\left(2, \lambda[i,k]\right).
    \label{eq:R_chi2_distribution}
\end{equation}
The non-centrality parameter $\lambda[i,k]$ corresponds to the power SNR for that specific bin, defined as:
\begin{equation}
    \lambda[i,k] \equiv \frac{2|\tilde{s}_i[k]|^2}{\langle|\tilde{n}_i[k]|^2\rangle}.
    \label{eq:lambda_ik_def}
\end{equation}
Note that the factor of 2 in \cref{eq:R_chi2_distribution} and \cref{eq:lambda_ik_def} arises from the properties of the complex Gaussian distribution governing the spectrum.

\subsection{Analytical framework}

Adopting the factorization strategy introduced in our previous work for \meth \cite{MyUnpublishedPaper}, we decompose the non-centrality parameter $\lambda[i,k]$ into two physically distinct components: a total power term and a normalized spectral shape term:
\begin{equation}
    \lambda[i,k] = \frac{2P_i}{\langle|\tilde{n}_i[k]|^2\rangle} \cdot \frac{|\tilde{s}_i[k]|^2}{P_i} \equiv \mathcal{L}_i \cdot \eta_i[k].
    \label{eq:lambda_factorization}
\end{equation}
The first factor, defined as the \textit{total power SNR statistic} $\mathcal{L}_i \equiv 2P_i / \langle|\tilde{n}_i[k]|^2\rangle$, quantifies the signal's average power within the $i$-th segment relative to the noise power in a single frequency bin. Here, $P_i$ represents the average power of the signal component after windowing, and a detailed discussion is in \cref{app:window_function_power}.

The second factor, $\eta_i[k] \equiv |\tilde{s}_i[k]|^2 / P_i$, is designated as the \textit{normalized power spectrum distribution factor}---or equivalently, the \textit{spectral leakage factor}. This normalized discrete function characterizes how the total signal power $P_i$ is partitioned among the discrete frequency bins $\{k\}$.  Crucially, $\eta_i[k]$ is intrinsic to the signal's frequency evolution and the windowing function, remaining invariant under changes in the noise background.

To systematically analyze the leakage pattern, it is advantageous to transition from the absolute frequency index $k$ to a signal-centric relative coordinate system. We define the instantaneous central frequency of the signal in dimensionless units as $k_{c,i} \equiv f_{c,i} T_\mathrm{DFT}$. This continuous frequency coordinate is decomposed into an integer component and a fractional offset:
\begin{itemize}
    \item The \textit{anchor bin}, denoted by $\lfloor k_{c,i} \rceil$, represents the integer DFT bin index closest to the true signal frequency.
    \item The \textit{fractional offset}, $o_{0,i} \equiv \lfloor k_{c,i} \rceil - k_{c,i}$, quantifies the displacement of the anchor bin center relative to the true frequency, bounded within the interval $o_{0,i} \in (-0.5, 0.5)$.
\end{itemize}
The relative frequency index $\kappa$ is then defined as the integer distance from this segment-specific anchor:
\begin{equation}
    \kappa \equiv k - \lfloor k_{c,i} \rceil.
\end{equation}
In this moving frame, $\kappa=0$ consistently refers to the peak (anchor) bin, while $\kappa=\pm 1$ denote the immediate sidebands. The effective offset for any relative bin $\kappa$ is given by $o_{\kappa,i} = \kappa + o_{0,i}$.

Consequently, the discrete leakage factors $\{\eta_i[\kappa]\}$ can be interpreted as equidistant samples drawn from an underlying continuous spectral leakage function, $\eta(o)$. The relationship is expressed as:
\begin{equation}
    \eta_i[\kappa] = \eta(o_{\kappa,i}) = \eta(\kappa + o_{0,i}).
    \label{eq:leakage_sampling}
\end{equation}
Here, the initial offset $o_{0,i}$ acts as a ``sampling phase'', dictating the discrete sampling grid along the continuous function. The temporal variation of the leakage pattern is thus entirely governed by the evolution of the sampling phase $\{o_{0,i}\}$, which is driven by the signal's intrinsic chirp.

\begin{figure}[h!]
    \centering
    \includegraphics[]{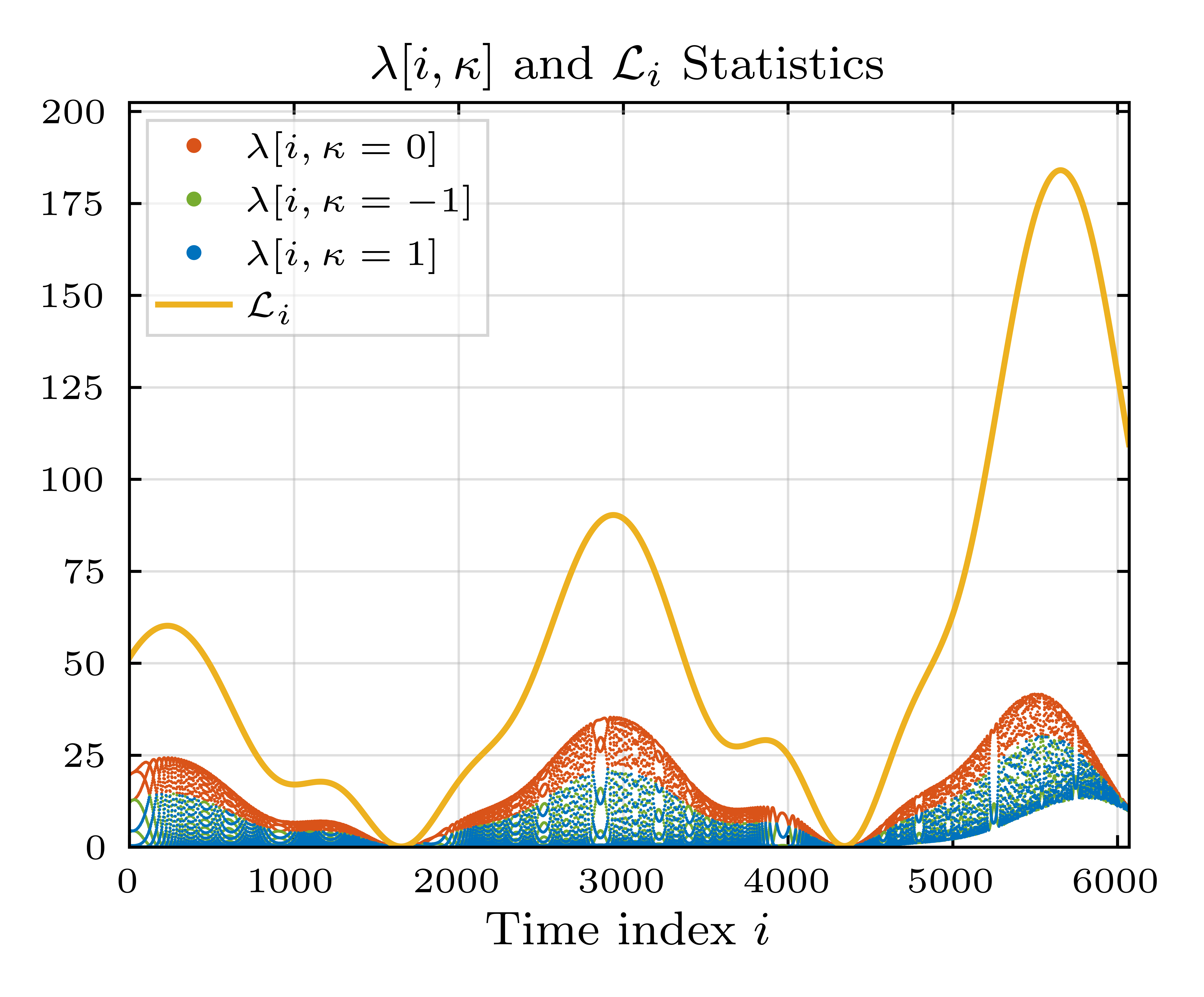}
    \includegraphics[]{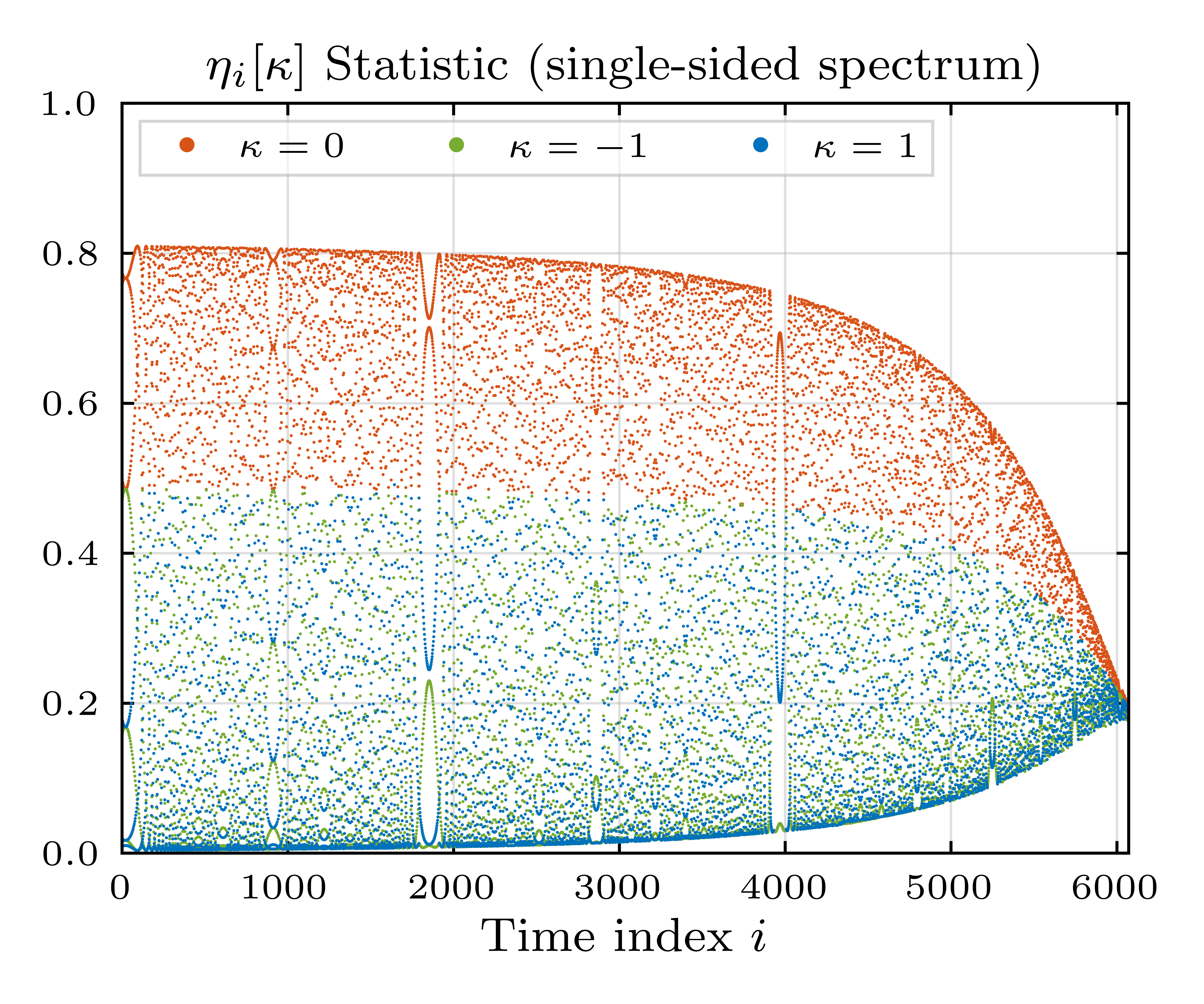}
    \caption{\textbf{Evolution of $\lambda[i,\kappa]$ and $\mathcal{L}_i$ (upper) and the spectral leakage factor $\eta_i[\kappa]$ (lower) over STFT time segments for a simulated mini-EMRI signal}. In both panels, the red dots represent the value in the anchoring frequency bin ($\kappa=0$), while the blue and green dots show the values in the adjacent bins ($\kappa=\pm1$). The yellow curve in the upper panel shows the total power statistic $\mathcal{L}_i$, calculated using a smoothed LIGO-H1 O3 PSD. The simulation uses a signal with $m_1 = 1.5 M_\odot$, $m_2 = 1 \times 10^{-5} M_\odot$, and $d = 8$ kpc, analyzed in the 100-200 Hz band. STFT parameters include $T_\mathrm{DFT} = 64$s, sampling frequency $f_s = 512$Hz, a Tukey window ($\alpha=0.5$), and 50\% overlap between segments.}
    \label{fig:stats}
\end{figure}

This decomposition is visualized in \cref{fig:stats}. For a direct comparison, we employ the identical signal configuration as in Fig. 8 of Ref. \cite{MyUnpublishedPaper}. However, a key distinction lies in the coherent integration time: guided by the analysis in \cref{fig:dmax_vs_T}, we extend the duration to $T=64\,\mathrm{s}$, relaxing the strict quasi-monochromatic limit ($T=8\,\mathrm{s}$) adopted in the previous work. The upper panel illustrates the temporal evolution of the total power statistic $\mathcal{L}_i$ (solid curve). Due to the extended integration time, the overall amplitude of $\mathcal{L}_i$ is significantly elevated compared to Ref.~[1], benefiting from the increased coherent gain. The curve's modulation reflects the secular evolution of the gravitational-wave amplitude coupled with the diurnal variation of the detector's antenna pattern. The scatter points depict the $\lambda$ values distributed across the anchor bin ($\kappa=0$, red) and its nearest neighbors ($\kappa=\pm1$, blue/green). A notable feature appears in the high-frequency regime (later times): while the total power remains high, the extended $T$ induces severe spectral leakage as the signal's chirp rate increases. Consequently, the power concentrated in individual bins (the dots) suffers greater loss relative to the total power. This phenomenon is quantified in the lower panel by the spectral leakage factor $\eta$, which exhibits rapid oscillatory behavior corresponding to the evolution of the sampling phase $o_{0,i}$.

The objective of the subsequent sections is to construct a precise analytical model for this universal leakage function $\eta(o)$.

\subsection{Validity of Approximations}
\label{sec:validity_of_approximations}

The extent to which a signal's frequency evolves within a single STFT segment is quantified by the dimensionless \textit{widening factor}, $w$. It is defined as the total frequency drift across the segment duration $T_\mathrm{DFT}$, normalized by the frequency resolution bin width $f_\mathrm{bin} = 1/T_\mathrm{DFT}$:
\begin{equation}
    w \equiv |\dot{f}| T_{\mathrm{DFT}}^2 = \frac{|\dot{f}|T_\mathrm{DFT}}{f_{\mathrm{bin}}}.
    \label{eq:widening_factor}
\end{equation}
This parameter serves as the governing metric for the validity of spectral leakage models $\eta(o)$. In the \textit{monochromatic limit} ($w \to 0$), where frequency evolution is negligible, the leakage function simplifies to the squared modulus of the window function's Fourier transform \cite{MyUnpublishedPaper}:
\begin{equation}
    \eta(o) = \left|\widetilde{W}(o)\right|^2.
    \label{eq:leakage_function_mono}
\end{equation}
Previous studies \cite{PhysRevD.70.082001,Miller:2018rbg} indicate that this approximation remains highly accurate for signals satisfying the \textit{quasi-monochromatic condition}, typically $w \lesssim 0.5$.

However, when $w$ is non-negligible, the monochromatic assumption breaks down. To rigorously assess the required model complexity, we analyze the signal phase $\Phi(t) = 2\pi \int f(t) dt$ by performing a Taylor series expansion of the instantaneous frequency $f(t)$ around the segment center. This expansion decomposes the phase into a polynomial in time: the linear term ($\propto t$) corresponds to the constant carrier frequency, the quadratic term ($\propto t^2$) arises from the linear frequency derivative $\dot{f}$ (linear chirp), and higher-order terms describe more complex frequency evolutions (e.g., $\ddot{f}$).

The validity of truncating this series is determined by the phase contribution of the neglected terms. We can define a critical widening factor $w_{\mathrm{crit},m}$, as the value of $w$ at which the contribution of the $m$-th order phase term becomes significant. Assuming the signal follows a local power-law evolution $\dot{f} = k f^n$, we derive a universal constant for the quadratic phase term—arising from the linear frequency drift $\dot{f}$—which is independent of specific signal parameters:
\begin{equation}
    w_{\mathrm{crit},2} = \frac{4}{\pi} \approx 1.2732.
    \label{eq:w_crit_constant}
\end{equation}
This value rigorously demarcates the boundary of the monochromatic approximation. When $w \ll w_{\mathrm{crit},2}$, the effects of linear frequency evolution are negligible. As $w$ approaches or exceeds this critical value, the quadratic phase term induces significant spectral spreading, necessitating a linear chirp model.

\begin{figure}[h!]
    \centering
    \includegraphics[]{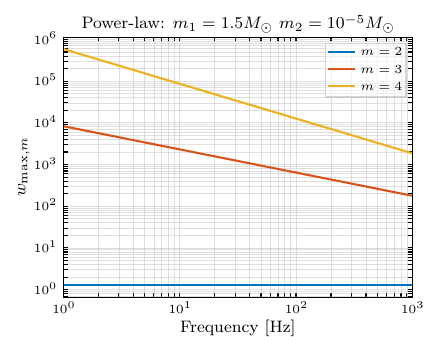}
    \includegraphics[]{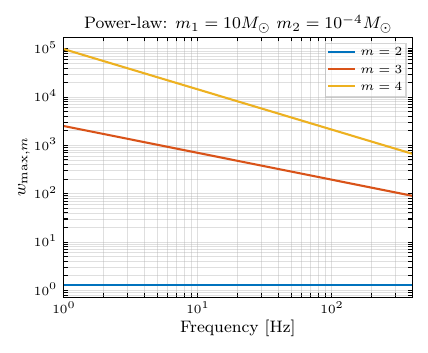}
    \caption{\textbf{Frequency evolution of the critical widening (see text) factors $w_{\mathrm{crit},m}$ (for orders $m=3, 4$), assuming a zeroth-order post-Newtonian inspiral model ($n=11/3$)}. The two panels correspond to different representative binary system parameters.}
    \label{fig:w_max_evolution}
\end{figure}

For phase terms of order $m > 2$ (arising from $\ddot{f}$ and higher derivatives), the critical thresholds $w_{\mathrm{crit},m}$ are not constant but depend on the specific signal parameters and frequency evolution. These are illustrated for two representative binary systems in \cref{fig:w_max_evolution}.

A key finding of this analysis is that the critical thresholds for higher-order terms are extremely high, with $w_{\mathrm{crit},3}$ typically ranging from $10^2$ to $10^5$. This has profound practical implications. To prevent excessive spectral smearing and preserve viable SNR, STFT-based analyses are typically constrained to a regime where $w \lesssim 10$. Since the practical limit on $w$ is orders of magnitude smaller than the critical values required for cubic phase terms to become relevant ($w \ll w_{\mathrm{crit},3}$), the effects of the second frequency derivative $\ddot{f}$---as well as all higher-order derivatives---can be safely neglected.

Consequently, this analysis rigorously justifies approximating the signal within each STFT segment as a \textit{linear chirp} (i.e., constant $\dot{f}$). This simplification underpins the entire analysis framework presented in this paper. A detailed derivation is provided in \cref{app:expansion}.

\subsection{Quantitative Description of Spectral Leakage}
\label{sec:leakage_description}

Having justified the linear chirp approximation in \cref{sec:validity_of_approximations}, we now provide a detailed mathematical formulation of the resulting spectral leakage pattern. This serves as a high-fidelity local model for the signal's spectral shape.

The distribution of signal power within an STFT segment is completely characterized by the continuous \textit{spectral leakage function}, $\eta(o,w)$. This function is parameterized by the dimensionless \textit{widening factor} $w$ and \textit{offset factor} $o$, which quantify the intra-segment frequency drift and the proximity to the central frequency, respectively.

\begin{figure}[h!]
    \centering
    \includegraphics[]{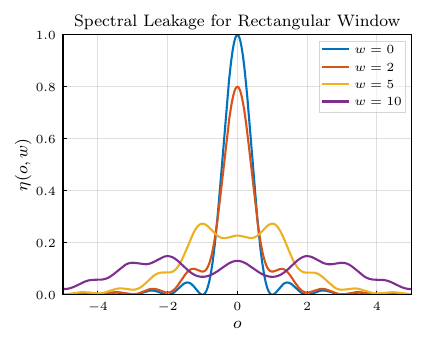}
    \includegraphics[]{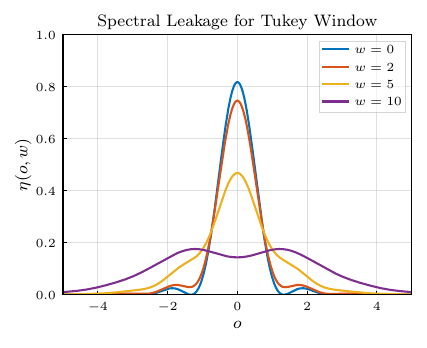}
    \caption{\textbf{The spectral leakage function $\eta(o, w)$ plotted against the dimensionless frequency offset $o$ for representative widening factors $w \in \{0, 2, 5, 10\}$}. The limiting case $w=0$ represents a monochromatic signal. The upper panel illustrates the leakage pattern for a Rectangular window, while the lower panel shows the pattern for a Tukey window.}
    \label{fig:eta_leakage}
\end{figure}

For a window function $W(\tau)$ defined on the normalized interval $\tau \in [-1/2, 1/2]$ and normalized to unit power (see \cref{app:window_function_power} for details), the leakage function has the general form:
\begin{equation}
    \eta(o,w) = \frac{1}{\pi w}\left|\int_{\sqrt{\pi}(o/\sqrt{w}-\sqrt{w}/2)}^{\sqrt{\pi}(o/\sqrt{w}+\sqrt{w}/2)} W\left(\frac{u}{\sqrt{\pi w}}-\frac{o}{w}\right) e^{i u^2} du\right|^2.
    \label{eq:leakage_function_general_maintext}
\end{equation}
\cref{fig:eta_leakage} illustrates representative profiles of the spectral leakage function. Generally, the function exhibits a decay as the magnitude of the offset factor $|o|$ increases, indicating that power is primarily localized around the carrier frequency. However, a crucial phenomenon is observed at higher widening: as $w$ increases, the signal power is smeared across a wider range of adjacent frequency bins.

This analytical model relies on the slow variation assumption: The signal's power is effectively constant over the segment duration $T_\mathrm{DFT}$. This is valid because the timescales of astrophysical amplitude evolution and detector antenna pattern modulation are typically much longer than the STFT coherence time.

Finally, for real-valued strain data, the power is distributed symmetrically between positive and negative frequencies. The leakage function derived above describes the one-sided spectrum; thus, the effective leakage for the positive frequency components is approximately half that of the complex equivalent derived in \cref{eq:leakage_function_general_maintext}. A rigorous treatment accounting for the real-valued nature of the signal is presented in \cref{app:linear_chirp}.

\section{Detection Statistic}
\label{sec:statistic}

\subsection{Statistical method}
\subsubsection{Distribution property}
The power ratio $R[i,k]$ in a pixel of the time-frequency map is a random variable that follows the probability density function (PDF):
\begin{equation}
    \mathrm{PDF}\left\{R=x| \lambda\right\} = e^{-x-\frac{\lambda}{2}} I_0\left(\sqrt{2\lambda x}\right), \quad x \ge 0,
\end{equation}
where $\lambda$ is the non-centrality parameter, representing the expected power SNR in that bin, and $I_0(\cdot)$ is the modified Bessel function of the first kind of order zero.

The mean and variance of $R[i,k]$ can be calculated directly from this distribution. For a given non-centrality parameter $\lambda[i,k]$, they are:
\begin{equation}
\begin{aligned}
    \mu_R[i,k] &=  1 + \frac{1}{2}\lambda[i,k], \\
    \sigma^2_R[i,k] & = 1 + \lambda[i,k].
\end{aligned}
\end{equation}
It is crucial to note that these are exact results and do not rely on any weak-signal approximations.




Conventional search methodologies, such as the widely implemented Hough transform pipeline \cite{PhysRevD.70.082001,Astone_2014,Miller:2018rbg,Miller:2024jpo,Menon:2025wce,Miller:2025ote,MyUnpublishedPaper}, typically employ a peak-finding strategy for data reduction. In this paradigm, the full information content of the normalized power spectrogram, $R[i,k]$, is condensed into a sparse binary ``peakmap'' by identifying bins that exceed a significance threshold and constitute local maxima along the frequency axis. This discretized representation subsequently serves as the input for track-finding algorithms to identify candidate signal trajectories.

In this work, we adopt a more direct approach by operating on the power ratio map itself, which retains more information and possesses well-defined statistical properties. For a given point in the parameter space (i.e., a specific set of signal parameters), there is a corresponding trajectory on the time-frequency map. Due to spectral leakage, the signal power is not confined to a single pixel per time segment but is distributed across several frequency bins. The bins directly on the trajectory are expected to have the highest SNR, with the SNR decreasing in adjacent bins as a function of their distance from the central track.



We propose the $\Sigma R$ statistic, computed by a weighted summation of the $R[i,k]$ values along a candidate trajectory. The weights $\omega_{i,k}$ are chosen to match the expected power SNR distribution with spectral leakage. 
To streamline the mathematical formulation, we define the \textit{generalized track set} $\mathcal{T}$ as the collection of time-frequency pixels $(i,k)$ associated with the signal. The detection statistic is then expressed as a unified weighted sum over this set:
\begin{equation}
    \Sigma R = \sum_{\mathcal{T}} \omega_{i,k} R[i,k].
    \label{eq:sum_R_statistic}
\end{equation}
Crucially, unlike the single-pixel tracks utilized in conventional methods, $\mathcal{T}$ can incorporate a cluster of frequency bins at each time step to recover the signal power dispersed by spectral leakage. Unless otherwise specified, we adopt a 3-pixel track width configuration (comprising the anchor bin and its two immediate sidebands, corresponding to relative indices $\kappa \in \{0, \pm 1\}$) as the default setting throughout this work.


Under the assumption that the power ratios $R[i,k]$ in different time-frequency pixels are statistically independent, the Central Limit Theorem implies that the aggregate statistic $\Sigma R$ follows an approximately Gaussian distribution. By the linearity of expectation and variance, its mean and variance are derived as the weighted sums of the individual moments:
\begin{equation}
\begin{aligned}
    \mu_{\Sigma R} &= \sum_{\mathcal{T}} \omega_{i,k}\left(1 + \frac{1}{2}\lambda[i,k]\right), \\
    \sigma^2_{\Sigma R} &= \sum_{\mathcal{T}} \omega_{i,k}^2\left(1 + \lambda[i,k]\right).
\end{aligned}
\label{eq:sum_R_mean_variance}
\end{equation}
It is important to acknowledge that in practice, the assumption of strict independence is not fully satisfied, primarily due to correlations induced by the overlap between time segments and spectral leakage between frequency bins. A detailed discussion on the impact of these correlations and the validity of this approximation is provided in \cref{app:correlations}.

To create a standardized detection statistic, we define the \textit{Critical Ratio} (CR). This quantity measures the significance of the observed $\Sigma R$ by normalizing it with respect to its expected value and standard deviation under the null hypothesis (i.e., in the absence of a signal, where $\lambda[i,k]=0$ for all pixels). The CR is therefore defined as the background-subtracted statistic divided by the background standard deviation:
\begin{equation}
    \mathrm{CR} \equiv \frac{\Sigma R - \sum_{\mathcal{T}} \omega_{i,k}}{\sqrt{\sum_{\mathcal{T}} \omega_{i,k}^2}}.
    \label{eq:CR_definition}
\end{equation}
By construction, the CR statistic for a track containing only noise is a random variable with a mean of 0 and a variance of 1. When a signal is present, the expected mean and variance of the CR statistic are shifted. Substituting the signal-dependent moments from \cref{eq:sum_R_mean_variance} into the definition yields:
\begin{equation}
\begin{aligned}
    \mu_\mathrm{CR} &= \frac{1}{2}\frac{\sum_{\mathcal{T}} \omega_{i,k}\lambda[i,k]}{\sqrt{\sum_{\mathcal{T}} \omega_{i,k}^2}}, \\
    \sigma^2_\mathrm{CR} &= 1 + \frac{\sum_{\mathcal{T}} \omega_{i,k}^2\lambda[i,k]}{\sum_{\mathcal{T}} \omega_{i,k}^2}.
\end{aligned}
\label{eq:CR_mean_variance}
\end{equation}
The quantity $\mu_\mathrm{CR}$ can be interpreted as the expected statistical significance of the detection for a given candidate track. The variance, $\sigma^2_\mathrm{CR}$, is close to unity for weak signals but increases with signal strength. The CR serves as the final, optimized statistic for identifying candidate signals in the data.

\subsubsection{Optimization of the weights}

The detection efficacy of the $\Sigma R$ statistic is critically dependent on the optimal selection of the weights $\omega_{i,k}$. Maximizing the detection significance is equivalent to maximizing the expected Critical Ratio, $\mu_\mathrm{CR}$. According to the Cauchy-Schwarz inequality, this optimum is achieved when the weights are directly proportional to the expected power SNR in each time-frequency pixel: $\omega_{i,k} \propto \lambda[i,k]$. This aligns with the matched filter principle and was also a key component of the winning solution in the Kaggle challenge \cite{Tenorio:2025vxv}: weights should be proportional to the expected signal strength. We note that since the CR statistic is invariant under global scaling transformations (i.e., $\omega_{i,k} \to c \cdot \omega_{i,k}$ for any $c>0$), the absolute normalization of the weights does not affect the detection significance.

In practice, however, constructing weights that are strictly proportional to the instantaneous $\lambda[i,k]$ is computationally intractable for blind searches. The difficulty arises from the sensitivity of the spectral leakage factor, $\eta_i[\kappa] = \eta(\kappa + o_{0,i})$, to the sub-bin sampling phase. As illustrated in the bottom panel of \cref{fig:stats}, $\eta$ oscillates rapidly with variations in the offset factor $o_{0,i}$. Consequently, predicting the exact value of $\eta_i[\kappa]$ for every pixel would require a priori knowledge of the source parameters with a precision far exceeding that of the search grid, which is unrealistic for wide-parameter-space surveys.

\begin{figure}[]
    \centering
    \includegraphics[]{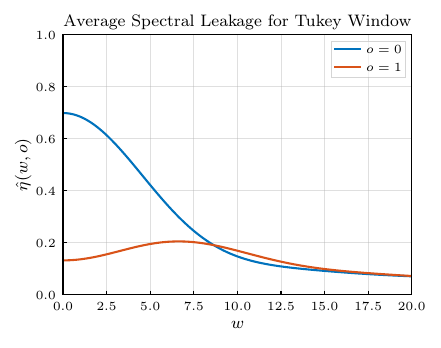}
    \includegraphics[]{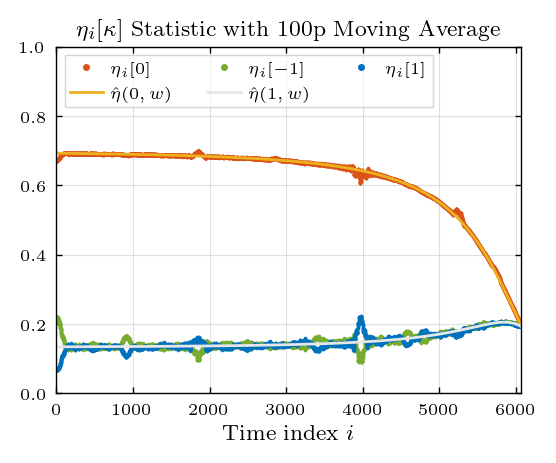}
    \caption{\textbf{The averaged spectral leakage function and its comparison with smoothed simulation data.}
    (Upper panel) The average spectral leakage function $\hat{\eta}(w,o)$. The blue curve shows the power fraction in the central bin ($\kappa=0$), while the red curve shows the power in the first adjacent bin ($\kappa=1$). As widening increases, power clearly leaks from the central bin into its neighbors.
    (Lower panel) A direct comparison validating the averaged model. The rapidly oscillating data from \cref{fig:stats} is smoothed using a 100-point moving average. This empirically smoothed result shows excellent agreement with the theoretical averaged leakage function, confirming its validity as a robust model for the expected power distribution.}
    \label{fig:average_sL}
\end{figure}

To circumvent this limitation, we implement a \textit{sampling phase-averaged strategy}, extending \meth from \cite{MyUnpublishedPaper}. Instead of relying on the highly sensitive instantaneous leakage, we define the \textit{averaged leakage function}, $\hat{\eta}(o, w)$, by integrating over a unit bin interval:
\begin{equation}
    \hat{\eta}(o,w) = \int_{o-1/2}^{o+1/2} \eta(o^\prime, w) \, do^\prime.
    \label{eq:averaged_leakage_integral}
\end{equation}
From this, we define the discrete \textit{averaged spectral leakage factor} for the $i$-th segment as $\hat{\eta}_i[\kappa] \equiv \hat{\eta}(\kappa, w_i)$. This formulation effectively marginalizes over the unknown sampling phase $o_{0,i}$, smoothing out the rapid oscillatory behavior of the instantaneous leakage. As demonstrated in \cref{fig:average_sL}, the averaged function captures the local mean spectral behavior over multiple segments, providing a deterministic template for weight construction.

Consequently, the practical weights are defined proportional to this averaged quantity:
\begin{equation}
    \omega_{i,k} \propto \mathcal{L}_i \hat{\eta}_i[k].
    \label{eq:practical_weights}
\end{equation}
Substituting these weights into \cref{eq:sum_R_mean_variance} and invoking the slow-variation approximations (see \cref{app:weights}), the expected mean and
variance of the CR simplify to:
\begin{equation}
\begin{aligned}
    \mu_{\mathrm{CR}} &= \frac{1}{2}\sqrt{\sum_{\mathcal{T}}\left(\mathcal{L}_i\hat{\eta}_i[k]\right)^2}, \\
    \sigma^2_{\mathrm{CR}} &= 1 + \frac{\sum_{\mathcal{T}}\left(\mathcal{L}_i\hat{\eta}_i[k]\right)^3}{\sum_{\mathcal{T}}\left(\mathcal{L}_i\hat{\eta}_i[k]\right)^2}.
\end{aligned}
\label{eq:CR_moments_practical}
\end{equation}
The adoption of this sampling phase-averaged weighting scheme yields a detection statistic that is significantly less sensitive to small parameter mismatches (specifically, errors in the sampling phase $o_{0,i}$). By effectively smoothing out the rapid fluctuations of the instantaneous value, this approach ensures that the statistic remains robust and computationally feasible for practical blind search campaigns.

\subsubsection{Robustness Against Non-Stationary Noise}

The statistical framework derived in the preceding sections relies on the idealized assumption that the detector noise is stationary and Gaussian, fully characterized by the spectral density $S_{n}(f)$. However, empirical data from interferometric detectors inevitably contain non-stationary artifacts---ranging from short-duration transients (``glitches'') to wandering spectral lines---that violate this premise. When such artifacts coincide with the generalized track set $\mathcal{T}$, they induce anomalously large power ratios ($R[i,k] \gg 1$), which act as outliers in the summation.

Since the $\Sigma R$ statistic is linear in power, these outliers can severely distort the background distribution under the null hypothesis. Specifically, they generate heavy non-Gaussian tails at high significance values, leading to an inflated false alarm rate (FAR) if left unmitigated.

To ensure the robustness of the search pipeline against these instrumental pathologies, we employ two standard mitigation strategies:
\begin{enumerate}
    \item \textbf{Power Ratio Truncation:} We impose a hard upper limit (saturation threshold) on the individual $R[i,k]$ terms contributing to the sum. This truncation prevents any single high-amplitude glitch from dominating the statistic $\Sigma R$, thereby suppressing the generation of spurious triggers.
    \item \textbf{Data Quality Vetoes:} Segments of data identified as contaminated by environmental or instrumental disturbances are flagged using standard Data Quality (DQ) definitions \cite{LIGO:2021ppb}. These flags are used to assign zero weight ($\omega_{i,k}=0$) to the corresponding time-frequency bins, effectively excising the compromised data from the accumulation process.
\end{enumerate}
The implementation of these cleaning procedures effectively restores the near-Gaussian behavior of the background noise, ensuring the reliability of the calculated detection confidence.

\subsection{Maximum detectable distance}

The primary objective of this statistical framework is to map the abstract detection statistic into physically interpretable constraints on the universe. Specifically, we aim to translate the statistical confidence thresholds---defined by the false alarm probability ($P_{\mathrm{fa}}$) and false dismissal probability ($P_{\mathrm{fd}}$)---into a metric of astrophysical sensitivity: the maximum distance, $d_{\mathrm{max}}$, at which a specific source can be confidently detected.

The detection performance is governed by two fundamental parameters: the false alarm probability $P_{\mathrm{fa}}$, which quantifies the risk of incorrectly classifying background noise as a signal, and the false dismissal probability $P_{\mathrm{fd}}$, which quantifies the risk of overlooking a genuine signal present in the data.

First, the choice of $P_{\mathrm{fa}}$ determines the detection threshold. Assuming the CR in pure noise follows a standard normal distribution (as justified in \cref{sec:statistic}), the detection threshold $\mathrm{CR}_{\mathrm{thr}}$ is given by the inverse complementary error function:
\begin{equation}
    \mathrm{CR}_{\mathrm{thr}} = \sqrt{2}~\mathrm{erfc}^{-1}(2P_{\mathrm{fa}}).
\end{equation}
Any candidate event with a CR value falling below this threshold is deemed statistically insignificant and rejected.

Conversely, to ensure a high probability of detecting a true signal, its expected CR value must lie sufficiently above this noise floor. For a given false dismissal probability $P_{\mathrm{fd}}$, the condition for the minimum required mean statistic is:
\begin{equation}
    \mu_{\mathrm{CR}} \geq \sqrt{2}~\sigma_{\mathrm{CR}}~\mathrm{erfc}^{-1}(2P_{\mathrm{fd}}) + \mathrm{CR}_{\mathrm{thr}}.
    \label{eq:min_mu_cr}
\end{equation}
This inequality establishes the minimum statistical significance required to claim a detection with the specified confidence level.

To translate this statistical requirement into a physical distance, we isolate the distance dependence within the total power statistic $\mathcal{L}_i$. We define the \textit{intrinsic strain amplitude}, $h_{e,i} \equiv h_{0,i} d$, which encapsulates the source's intrinsic evolution independent of its distance $d$. Consequently, the total power SNR statistic can be factored as:
\begin{equation}
    \mathcal{L}_i = \frac{1}{d^2} \left(2 h_{e,i}^2 Q_i^2  \frac{T_\mathrm{DFT}}{S_{n,i}} \right),
\end{equation}
where $Q_i$ represents the antenna pattern response and $S_{n,i}$ is the noise power spectral density. A detailed discussion is in \cref{app:window_function_power}.

As detailed in \cref{app:dmax}, when estimating the detectable distance, we are by definition operating in the \textit{weak-signal limit}. In this regime, the signal's contribution to the variance is negligible compared to the noise background, implying $\sigma_{\mathrm{CR}}^2 \approx 1$. Applying this approximation allows us to decouple the distance dependence and derive the central result of our sensitivity analysis:
\begin{widetext}
\begin{equation}
    d_{\mathrm{max}} = \left[ \sqrt{2}~\mathrm{erfc}^{-1}(2P_{\mathrm{fa}}) + \sqrt{2}~\mathrm{erfc}^{-1}(2P_{\mathrm{fd}}) \right]^{-1/2}  \left[ \frac{1}{4}\sum_{\mathcal{T}}h_{e,i}^4Q_i^4\frac{T_\mathrm{DFT}^2}{S_{n,i}^2}\hat{\eta}_i^2[\kappa]\right]^{1/4}.
\label{eq:d_max}
\end{equation}
\end{widetext}
This analytical formula provides a direct mapping from source parameters to the detection horizon. For any given astrophysical model, it enables the calculation of the maximum sensitive distance, thereby quantifying the effective volume of the universe accessible to our search.

\subsection{Effective parameter sphere}
\label{sec:mismatch_framework}

Practical searches for gravitational-wave signals in noisy data necessitate the use of a discrete grid of waveforms, known as a \textit{template bank}, to span the vast parameter space of potential sources. Since the parameters of astrophysical signals are continuous variables, a real signal will rarely align perfectly with any single discrete template. Consequently, it is imperative to quantify the degradation of the detection statistic caused by the parameter mismatch between the search template and the true signal.

To formalize this, we define the output of the search pipeline as the expected Critical Ratio, $\mu_\mathrm{CR}(\mathcal{P}_t|\mathcal{P}_r)$, obtained when probing for a signal with true parameters $\mathcal{P}_r$ using a template with parameters $\mathcal{P}_t$. While the analytical form of $\mu_\mathrm{CR}$ is derived in \cref{app:weights}, its value is fundamentally governed by the degree of overlap between the template and the signal waveforms.

To quantify the recovery efficiency, following the formalism in \cite{MyUnpublishedPaper}, we define the \textit{fitting factor} (FF). This metric represents the ratio of the expected statistic recovered by the template to the maximum achievable statistic if the template were perfectly matched to the signal:
\begin{equation}
    \mathrm{FF}(\mathcal{P}_t|\mathcal{P}_r) \equiv \frac{\mu_\mathrm{CR}(\mathcal{P}_t|\mathcal{P}_r)}{\mu_\mathrm{CR}(\mathcal{P}_r|\mathcal{P}_r)}.
    \label{eq:fitting_factor}
\end{equation}
By definition, the fitting factor is a normalized measure of effectualness, ranging from 0 (no match) to 1 (perfect match).

Building on this, we define the \textit{mismatch} (MM) to represent the fractional loss in astrophysical sensitivity. Unlike standard conventions which often define mismatch in terms of power SNR loss, we adopt a definition directly tied to the cosmological reach. As established in \cref{eq:d_max} and \cref{eq:CR_moments_practical}, the detection distance as $d_{\mathrm{max}} \propto \sqrt{\mu_\mathrm{CR}}$. Consequently, the fractional loss in detectable distance is given by $1 - \sqrt{\mathrm{FF}}$. We therefore define the mismatch as:
\begin{equation}
    \mathrm{MM} \equiv 1 - \sqrt{\mathrm{FF}}.
    \label{eq:mismatch}
\end{equation}
This physically motivated definition renders the mismatch an exceptionally intuitive metric: a value of $\mathrm{MM} = 0.01$ corresponds to a 1\% reduction in the detector's reach distance.

This metric serves as the foundational criterion for template bank construction. The objective is to populate the parameter space with templates such that the mismatch for any arbitrary signal remains below a predefined tolerance, $\mathrm{MM}_{\mathrm{max}}$, thereby ensuring uniform sensitivity across the target parameter space. This requirement can be formulated as a classic geometric covering problem. For a given template, the condition $\mathrm{MM} \leq \mathrm{MM}_{\mathrm{max}}$ defines a high-dimensional volume around the template center within which signals are detected with sufficient efficiency. In the limit of small parameter offsets, this iso-mismatch surface is well-approximated by a hyper-ellipsoid, often referred to as the \textit{metric ellipse} or \textit{effective detection sphere}. The task of constructing a template bank is thus transformed into the challenge of efficiently tiling the parameter space with these overlapping ellipsoids. The overarching goal is to minimize the number of templates to reduce computational cost while guaranteeing that no "holes" exist where the sensitivity drops below the required threshold. The specific algorithms implementing this stochastic geometric placement are detailed in subsequent work.

\section{Results}
\label{sec:result}
In this section, we apply the comprehensive theoretical framework developed in the preceding sections to a practical search scenario. We leverage our description of spectral leakage, which is valid beyond the traditional quasi-monochromatic approximation, to guide the optimization of key search parameters. By employing the new $\Sigma R$ statistic, we demonstrate how this theoretically-grounded approach enables a significant enhancement in search sensitivity. The ultimate metric for this sensitivity is the effective detection volume, $V_\mathrm{eff}$, which represents the astrophysical reach of our search averaged over all source sky locations and orientations, as formally defined in \cref{eq:Veff_full_integral} of \cref{app:Veff}. This volume is computed via the effective distance, $d_\mathrm{eff}$. The calculation is made tractable by incorporating the fully-averaged antenna-pattern factor $\langle Q \rangle$ (from \cref{eq:Q_averaged}) which encapsulates the complete geometric average over sky position, orientation, and sidereal time. 

We now apply our framework to a concrete target: a fiducial mini-EMRI system consisting of a typical neutron star ($m_1=1.5 M_\odot$) and a strange compact object ($m_2=10^{-5} M_\odot$). Our analysis focuses on the 100-200 Hz frequency band, where the signal is expected to be most prominent for ground-based detectors. \cref{fig:stats} illustrates the key statistics for a simulated signal from such a system.

Our primary goal is to determine the optimal coherent time $T_{\mathrm{DFT}}$ that maximizes search sensitivity. Previous searches were restricted by the quasi-monochromatic approximation, which strictly limited the coherent time to $T_{\mathrm{DFT}}\leq 8\ \mathrm{s}$ to avoid excessive spectral leakage. Our new framework, which analytically models and accounts for this leakage, allows us to break free of this long-standing constraint. This enables a comprehensive optimization of $T_\mathrm{DFT}$ over a wide range to maximize search sensitivity.

\begin{figure}[]
    \centering
    \includegraphics[]{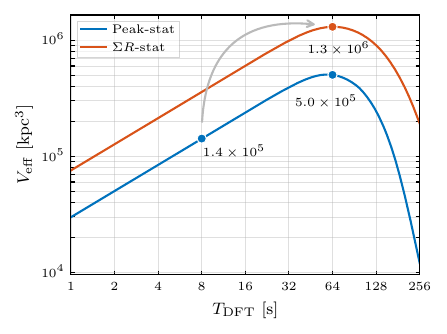}
    \caption{\textbf{The effective detection volume $V_\mathrm{eff}$ as a function of coherent integration time, $T_\mathrm{DFT}$, for the fiducial mini-EMRI system, computed using the O3 PSD for the LIGO Hanford detector}. The performance is evaluated for a set $P_\mathrm{fd}=5\%$ and $P_\mathrm{fa}=1\%$. STFT uses the Tukey window function ($\alpha=0.5$) with a 50\% overlap. The red curve illustrates the performance of our novel $\Sigma R$ statistic. In comparison, the blue curve depicts the performance of a statistic based on the Hough transform strategy. Instead of a full pipeline execution, this peak statistic is computed by summing binary pixels along the signal trajectory, mimicking the standard Hough approach which selects peaks based on a power ratio threshold (here $\theta=2.5$) and a spectral local maximum condition.
    The performance at $T=8$s, the optimal choice under the quasi-monochromatic assumption, is marked in the panel. Our new method, which transcends this limitation, finds a true optimal coherent time at $T=64$s.}
    \label{fig:dmax_vs_T}
\end{figure}

\cref{fig:dmax_vs_T} illustrates the direct consequence of this advancement, plotting the effective detection volume as a function of $T_\mathrm{DFT}$ for both our new method and a conventional approach. For both pipelines, the detection volume initially rises with longer integration times—a direct result of the coherent gain in SNR (see \cref{app:dmax}). However, as $T_\mathrm{DFT}$ continues to increase, the signal's rapid frequency evolution begins to spread its power across many frequency bins. This spectral leakage eventually dominates, causing the sensitivity to decline. While the conventional method is confined to the quasi-monochromatic limit at $T_\mathrm{DFT}=8\,$s, our method identifies a true global optimum at $T_\mathrm{DFT}=64\,$s. At this optimal integration time, our method yields an approximately 10-fold increase in the effective detection volume compared to the previous state-of-the-art. This represents a dramatic, order-of-magnitude improvement in search sensitivity.

To demonstrate the practical impact of our method in a \textit{realistic broadband search scenario}, we move beyond narrow-band analyses and consider a full signal trajectory evolving from $30\,\mathrm{Hz}$ to $1000\,\mathrm{Hz}$ over an observation period of 66 days. In this context, we allow both the conventional and the proposed pipelines to freely optimize their parameters to achieve their respective maximum performance.

For the conventional Hough-transform pipeline, we perform a global optimization to select the single best coherent time $T_{\mathrm{DFT}}$ that maximizes the total detection volume over this wide band. The optimization yields a best-case scheme of $T_{\mathrm{DFT}}=32\,\mathrm{s}$. Although this is the mathematical optimum for the standard pipeline, it effectively restricts the sensitive search window to a narrow frequency band of $f \in [80, 135]\,\mathrm{Hz}$—sacrificing the vast majority of the signal's time-frequency evolution.

In stark contrast, the $\Sigma R$ framework is intrinsically capable of local optimization. Leveraging our rigorous leakage model, we are free to implement a frequency-layered strategy that adapts to the signal's evolution across the entire 30--1000\,Hz band, which extends a framework presented for searches for compact binary inspirals in future ground-based GW interferometers \cite{Tenorio:2025gci} to mini-EMRIs. By partitioning the bandwidth into distinct layers and optimizing the coherent time locally, we maintain peak sensitivity throughout the signal's lifetime.

\begin{figure}[]
    \centering
    \includegraphics[]{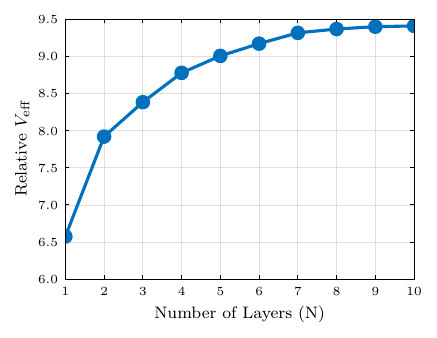}
    \caption{\textbf{Relative effective detection volume as a function of the number of frequency layers $N$}. The $y$-axis represents the ratio of the effective detection volume to that of the reference case, which is the optimal configuration of the conventional Hough-transform pipeline.}
    \label{fig:freq_layers}
\end{figure}

The result of this comparison, illustrated in \cref{fig:freq_layers}, reveals a twofold advantage of our framework. First, even in the simplest single-layer configuration ($N=1$), the $\Sigma R$ method already yields a $6.5$-fold increase in detection volume compared to the optimized Hough baseline. Second, as we exploit the freedom to implement the frequency-layered strategy, the performance improves further. As the number of layers increases from $N=1$ to $N=10$, the improved matching of the coherent time to the local signal evolution compounds the gain, ultimately achieving an effective detection volume that is approximately \textit{an order of magnitude larger} than the conventional limit.

\begin{figure}[h!]
    \centering
    \includegraphics[]{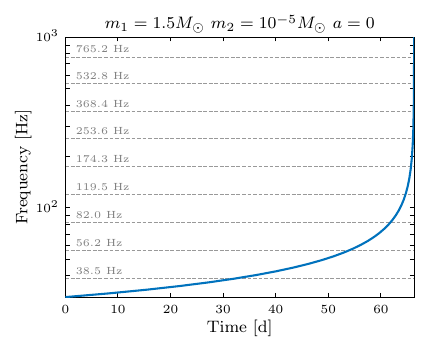}
    \caption{\textbf{Visualization of the optimal frequency-layering strategy for the target mini-EMRI signal}. The blue curve represents the signal's frequency evolution over time. The entire search band is partitioned into $N=10$ distinct layers by critical frequencies (horizontal dashed lines), with the coherent time $T_\mathrm{DFT}$ decreasing stepwise from $1024\,\mathrm{s}$ in the lowest layer to $2\,\mathrm{s}$ in the highest.}
\label{fig:layer_intervals}
\end{figure}

To elucidate the practical implementation of the multi-layer strategy, \cref{fig:layer_intervals} presents the detailed configuration for the representative case of $N=10$ layers. The plot visualizes the target signal's time-frequency trajectory overlaid with the optimized layer boundaries, defined by a set of critical frequencies. Details about the frequency layering strategy are discussed in \cref{app:Freq_layer}.

Enabled by this frequency-layered architecture, the coherent integration time $T_\mathrm{DFT}$ is dynamically adapted to the signal's local chirp rate, decreasing stepwise from a maximum of $1024\,\mathrm{s}$ in the lowest frequency layer to $2\,\mathrm{s}$ in the highest layer. This stepwise reduction is critical to counterbalance the rapidly accelerating frequency evolution of the inspiral. Consequently, the widening factor remains constrained within the optimal regime throughout the entire observation. By preventing excessive spectral leakage at high frequencies while preserving long coherent integration times at low frequencies, this configuration realizes the order-of-magnitude expansion in detection volume quantified in \cref{fig:freq_layers}.



\section{Conclusions}

In this paper, we have taken a major step forward in developing a search pipeline for mini-EMRI signals. We have extended our method, \texttt{\textSigma Track}, to handle the case of significant spectral leakage via a new statistic, $\Sigma R$, that allows us to significantly increase the analysis coherence time with respect to that used in standard methods that only consider the signal as monochroamtic within each fast Fourier transform and do not consider the impact of spectral leakage. additionally, we have derived a new sensitivity estimate of our method, and shown that by accurately modeling spectral leakage and optimizing the coherent time, \texttt{\textSigma Track} boosts the effective detection volume for this system by an order of magnitude. This dramatic enhancement in search sensitivity translates directly into a correspondingly higher probability of detecting these elusive mini-EMRI sources with current and future gravitational-wave observatories.

In future work, we plan to generalize our method further to handle non-stationary, non-Gaussian noise, which will be essenetial to apply \texttt{\textSigma Track} to real data. Moreover, following \cite{MyUnpublishedPaper}, we will determine a discretization of the search parameter space based on a mismatch between signals in the time-frequency plane to generate a bank of mini-EMRI templates to ensure that a real search is tractable.
\begin{acknowledgments}


This material is based upon work supported by NSF's LIGO Laboratory which is a major facility fully funded by the National Science Foundation.

We acknowledge the Python scientific ecosystem, particularly \texttt{NumPy} \cite{harris2020array}, \texttt{SciPy} \cite{2020SciPy-NMeth}, and \texttt{Matplotlib} \cite{Hunter:2007}, which were used in the analysis and visualization.
This work is supported by the National Natural Science Foundation of China (NSFC) under Grant No. 12347103 and 12547104.
\end{acknowledgments}

\appendix

\section{Phase Expansion in the Fourier Domain}
\label{app:expansion}

In this appendix, we provide a detailed derivation of the conditions under which a general frequency-evolving signal can be approximated by either a monochromatic or a linear-chirp model within a single coherent integration time, $T$. This analysis forms the mathematical basis for extending the coherent integration time beyond the traditional quasi-monochromatic limit.

We begin with a general, continuous frequency-modulated signal, which can be expressed in complex form as:
\begin{equation}
    s(t)=A(t)e^{i\phi(t)},
\end{equation}
where $A(t)$ is the time-varying amplitude and $\phi(t)$ is the phase. The phase is determined by the integral of the signal's instantaneous frequency $f(t)$:
\begin{equation}
    \phi(t) = \phi_0 + 2\pi\int_0^t f(\tau) d\tau,
\end{equation}
where $\phi_0$ is the initial phase at $t=0$.

To analyze this signal in the frequency domain, we consider its STFT of duration $T$ centered at $t=0$. The Fourier transform of the windowed segment is:
\begin{equation}
\begin{aligned}
    \tilde{s}(f)
    &=\int_{-T/2}^{T/2} A(t) \exp\left(i\phi_0+i2\pi\int_0^t f(\tau) d\tau\right) e^{-i 2 \pi f t} dt.
\end{aligned}
\end{equation}
To make this integral tractable, we Taylor-expand the instantaneous frequency $f(t)$ around the center of the window, $t=0$:
\begin{equation}
    f(t) = f_0 + \dot{f}_0 t + \frac{1}{2}\ddot{f}_0 t^2 + \cdots,
\end{equation}
where $f_0, \dot{f}_0, \ddot{f}_0$ are the frequency and its time derivatives evaluated at $t=0$. Integrating this series gives the full phase expansion:
\begin{equation}
\begin{aligned}
    \phi(t) &= \phi_0 + 2\pi\int_0^{t} \left( f_0 + \dot{f}_0 \tau + \frac{1}{2} \ddot{f}_0 \tau^2 + \cdots\right) d\tau\\
    &= \phi_0 + 2\pi\left( f_0 t + \frac{1}{2}\dot{f}_0 t^2 + \frac{1}{6}\ddot{f}_0 t^3 + \cdots\right).
\end{aligned}
\end{equation}
For many astrophysical signals, the amplitude $A(t)$ evolves on a much longer timescale than the coherent integration time $T$. We can therefore make the standard assumption that the amplitude is constant within the window, $A(t) \approx A(0)$. This allows us to simplify the Fourier integral significantly:
\begin{equation}
\begin{aligned}
    \tilde{s}(f) \approx& A(0)\int_{-T/2}^{T/2}e^{i\phi(t)} e^{-i 2 \pi f t} dt \\
    =& A(0) \int_{-T/2}^{T/2} e^{\left(i\phi_0 + i 2 \pi f_0 t + i \pi \dot{f}_0 t^2 + \frac{i\pi}{3}\ddot{f}_0t^3+\cdots\right)}e^{-i2\pi ft} dt.
\end{aligned}
\end{equation}
The central challenge in simplifying the integral is to determine at which order the phase expansion can be truncated. We can truncate the series after the quadratic term ($t^2$) if the combined phase contribution from all higher-order terms (cubic and beyond) is much less than one radian over the entire integration interval. When this condition, $|\sum_{j=3}^{\infty} 2\pi \frac{f_0^{\{j-1\}}}{j!}t^j| \ll 1$, is met, we can use the approximation $e^x = 1+x+x^2/2+\cdots \approx 1$ for the higher-order part of the exponent.

This allows us to separate the integral into a primary component and a small residual term, which can then be neglected:
\begin{equation}
\begin{aligned}
    &\int_{-T/2}^{T/2} e^{\left(i\phi_0 + i 2 \pi f_0 t + i \pi \dot{f}_0 t^2 + \frac{i}{3}\pi\ddot{f}_0t^3+\cdots\right)} e^{-i 2 \pi f t} dt\\
    =&\int_{-T/2}^{T/2} e^{\left(i\phi_0 + i 2 \pi f_0 t + i \pi \dot{f}_0 t^2 \right)}\left(1+ \frac{i}{3}\pi\ddot{f}_0t^3+\cdots\right) e^{-i 2 \pi f t} dt\\
    \approx&\int_{-T/2}^{T/2} e^{\left(i\phi_0 + i 2 \pi f_0 t + i \pi \dot{f}_0 t^2 \right)} e^{-i 2 \pi f t} dt.
\end{aligned}
\label{eq:integral_term}
\end{equation}
To formalize this truncation, we must quantify the contribution of each higher-order term. We first adopt a general power-law signal model, $\dot{f}=kf^n$, which is representative of many astrophysical sources. Under this model, the $m$-th derivative of the frequency can be written in a compact form:
\begin{equation}
    f^{\{m\}}_0=\left.\frac{d^{m}f(t)}{dt^m}\right|_{t=0} =C(m,n)k^{m} f_0^{m(n-1)+1},
\end{equation}
where the coefficient is $C(m,n)=\prod_{j=0}^{m-1} \left[j(n-1)+1\right]$. The phase contribution corresponding to the $t^m$ term in the expansion is:
\begin{equation}
\begin{aligned}
    \text{PT}(m) &= 2\pi \int_0^t \frac{f^{\{m-1\}}_0}{(m-1)!}\tau^{m-1}d\tau\\
    &= 2\pi \frac{f^{\{m-1\}}_0}{m!}t^m \\
    &= 2\pi\frac{ C(m-1,n)}{m!} k^{m-1} f_0^{(m-1)(n-1)+1} t^m.
\end{aligned}
\label{eq:term_m}
\end{equation}
The influence of this term is maximal at the edges of the window, $t=\pm T/2$. To create a dimensionless metric, we introduce the widening factor $w \equiv |\dot{f}_0|T^2$. By substituting $t=T/2$ into \cref{eq:term_m} and re-expressing $T$ in terms of $w$, we find the maximum phase contribution of the $m$-th order term:
\begin{equation}
\begin{aligned}
    \max|\text{PT}(m)| &= 2\pi\frac{ C(m-1,n)}{m!} k^{m-1} f_0^{(m-1)(n-1)+1} \left(\frac{T}{2}\right)^m\\
    &= \frac{2\pi C(m-1,n)}{m!2^m} k^{\frac{m}{2}-1} f_0^{\frac{mn}{2}-m-n+2} w^{\frac{m}{2}}.
\end{aligned}
\end{equation}
We now define a critical widening factor, $w_{\mathrm{crit}, m}$, as the value of $w$ for which this maximum phase contribution equals 1 radian. Setting the above expression to 1 and solving for $w$ yields:
\begin{equation}
    w_{\mathrm{crit},m} = \left(\frac{m! 2^m}{2\pi C(m-1,n)} k^{1-\frac{m}{2}} f_0^{m+n-\frac{mn}{2}-2}\right)^{\frac{2}{m}}.
    \label{eq:w_max}
\end{equation}
This powerful result allows us to express the maximum phase contribution of any term in a very elegant form:
\begin{equation}
    \max|\text{PT}(m)| = \left(\frac{w}{w_{\mathrm{crit},m}}\right)^{\frac{m}{2}}.
    \label{eq:phase_contribution_ratio}
\end{equation}
This inequality provides a clear condition for truncating the phase expansion. For a given order $m$, if the signal's widening factor $w$ is much smaller than the critical factor $w_{\mathrm{crit},m}$, then the $m$-th order phase term is negligible. It is important to note that this linear approximation is only meaningful when the ratio in \cref{eq:phase_contribution_ratio} is small. When the phase contribution is of order $2\pi$ or larger, its periodic nature means its absolute value is no longer a good measure of its impact on the integral.

The framework provides a explanation for the validity of commonly used signal approximations. We summarize the key conclusions for the two most important cases.

\begin{itemize}
    \item \textbf{The Quasi-Monochromatic Approximation} ($m=2$): The phase expansion begins with the constant ($m=0$) and linear ($m=1$) terms, which represent the signal's central time frequency. The first term describing frequency evolution is the quadratic term ($m=2$). The critical widening factor for this term is a universal constant for any power-law signal: $w_{\mathrm{max},2} = 4/\pi \approx 1.2732$. This result provides the formal justification for the widely-used quasi-monochromatic condition. To neglect all frequency evolution, the signal's widening factor $w$ must be much less than $4/\pi$. The practical choice of $w < 0.5$ corresponds to the condition $w \ll w_{\mathrm{max},2}$, ensuring the signal can be accurately treated as monochromatic over the time duration $T$.
    \item \textbf{The Linear-Chirp Approximation} ($m=3$): When the quasi-monochromatic condition is violated ($w \gtrsim 1$), we must include the quadratic phase term. The next question is whether the cubic term ($m=3$) is also necessary. As shown in the examples in \cref{fig:w_max_evolution}, the critical widening factor for the cubic term, $w_{\mathrm{max},3}$, is typically very large for astrophysical systems, often on the order of $10^3-10^4$.
\end{itemize}

This has a crucial implication: for any practical search where the widening factor is in the range of $0 < w \lesssim 10$, the condition $w \ll w_{\mathrm{max},3}$ is strongly satisfied. This means that the cubic and all higher-order phase terms can be safely ignored. Therefore, even when a signal evolves rapidly, a linear-chirp model (retaining terms up to $m=2$) provides an extremely accurate representation of the signal's phase. This is the fundamental principle that enables our search method to extend the coherent integration time far beyond traditional limits.

\section{Linear chirp signal model}
\label{app:linear_chirp}

We begin with the mathematical model for a discrete, complex linear-chirp signal sampled $N$ times over a total duration $T$. The time-domain sequence is written as:
\begin{equation}
    s[n] = \exp\left(i\left[\phi_0 + 2\pi f_0\frac{ T}{N}n + \pi \dot{f}\frac{ T^2}{N^2} n^2\right]\right),
\end{equation}
where $\phi_0$ is the initial phase at the start of the segment. The parameter $f_0$ represents the instantaneous frequency at $t=0$ (the start of the interval), and $\dot{f}$ is the constant chirp rate. For this derivation, we assume $\dot{f} > 0$, such that the instantaneous frequency evolves as $f(t)=f_0+\dot{f}t$ over the segment. 

It is important to clarify why we specifically analyze a complex signal with unit amplitude, which can ensure that the time-domain power is constant at unit. This convention is analytically convenient because from the definition \cref{eq:lambda_factorization} the power spectrum value in a single bin $|\tilde{x}[k]|^2$, directly corresponds to our spectral leakage function, $\eta(w,o)$.

The DFT of this signal at frequency bin $k$ is given by:
\begin{equation}
\begin{aligned}
    \tilde{s}[k]
    &= \frac{1}{N}\sum_{n=0}^{N-1} s[n]\, e^{-i 2\pi \frac{k}{N} n} \\
    &= \frac{e^{i\phi_0}}{N}\sum_{n=0}^{N-1} \exp\left\{ i\left[ 2\pi\left(f_0 T-k\right) \frac{n}{N} + \pi\dot{f}T^2\frac{n^2}{N^2}  \right]\right\}.
\end{aligned}
\end{equation}

For a sufficiently large number of samples $N$, where the normalized time variable $\mu=n/N$ becomes quasi-continuous, this discrete sum can be accurately approximated by its corresponding integral. By making the substitution $1/N\sum_{n=0}^{N-1} \to \int_0^1 d\mu$, we obtain:
\begin{equation}
\begin{aligned}
    \tilde{s}[k]
    &\approx e^{i\phi_0}\int_{0}^{1} \exp\left\{ i\left[ 2\pi\left(f_0 T-k\right)\mu + \pi\dot{f}T^2\mu^2 \right] \right\}d\mu.
\end{aligned}
\end{equation}

This integral is of the Fresnel type. By completing the square in the exponent and performing the standard change of integration variable, the result can be expressed in terms of the complex Fresnel integral:
\begin{equation}
    \tilde{s}[k]
    = \frac{1}{\sqrt{\pi \dot{f}T^2}}e^{i\phi_0-iv^2}
    \int_{v}^{v + \sqrt{\pi\dot{f}T^2}} e^{i u^2}\, du,
\end{equation}
where the lower integration limit $v$ is a dimensionless quantity defined as:
\begin{equation}
    v \equiv \sqrt{\pi}\,\frac{f_0T-k}{\sqrt{\dot f\,T^2}}.
\end{equation}

To generalize the analysis and make the spectral behavior more transparent, we introduce two convenient dimensionless quantities:
\begin{itemize}
    \item The \textit{widening factor}, $w$, which represents the dimensionless chirp strength over the segment duration:
    \begin{equation}
        w \equiv \dot{f} T^2.
    \end{equation}
    \item The \textit{offset factor}, $o$, which measures the frequency offset (in units of bins) between the bin frequency $f_k=k/T$ and the signal's central time frequency over the segment, $f_c = f_0 + \dot{f}T/2$:
    \begin{equation}
    \begin{aligned}
        o &\equiv (f_k-f_c)T \\
        &= \left(\frac{k}{T}-f_0-\frac{1}{2}\dot{f}T\right)T\\
        &=(k-f_0T)-\frac{1}{2}w.
    \end{aligned}
    \end{equation}
\end{itemize}

Expressed in terms of these dimensionless variables, the normalized power in a single frequency bin $k$, which we denote as the spectral leakage factor $\eta(w,o)$, is given by the squared modulus of the DFT coefficient:
\begin{equation}
\begin{aligned}
    \eta(o,w) \equiv \frac{1}{\pi w}\left|\int_{\sqrt{\pi}(o/\sqrt{w}-\sqrt{w}/2)}^{\sqrt{\pi}(o/\sqrt{w}+\sqrt{w}/2)}e^{i u^2}d u\right|^2.
\end{aligned}
\label{eq:eta_final_rewrite}
\end{equation}
For the case of a negative chirp rate ($\dot{f}<0$), the result for $\eta(w,o)$ remains identical due to the symmetries of the complex exponential and the squared magnitude operation. The phase of $\tilde{s}[k]$ would change, but the power spectrum is unaffected.

It is crucial to recognize that the function $\eta(o,w)$ describes a continuous power spectrum as a function of the offset $o$. The power spectrum calculated from a DFT $|\tilde{s}[k]|^2$, is a discrete sampling of this continuous profile, evaluated at the specific offset values corresponding to each integer frequency bin $k$.

The preceding analysis was based on an implicit rectangular window, which has a constant value over the integration time $T$. However, in practical signal processing, other window functions (such as Hann, Hamming, or Tukey) are often applied to the time-domain data. It is therefore crucial to generalize our model for an arbitrary, power-normalized window function.

Let $W(\tau)$ be a window function defined over the normalized time interval $\tau \in [-1/2, 1/2]$, satisfying the power-normalization condition $\int_{-1/2}^{1/2} |W(\tau)|^2 d\tau = 1$. When this window is applied to the time-domain signal, the derivation for the spectral leakage factor $\eta$ follows the same procedure, but with the window function carried through the integral.

After performing the same change of variables that leads to the Fresnel integral representation, the generalized spectral leakage function takes the form:
\begin{equation}
    \eta(o,w) = \frac{1}{\pi w}\left|\int_{\sqrt{\pi}(o/\sqrt{w}-\sqrt{w}/2)}^{\sqrt{\pi}(o/\sqrt{w}+\sqrt{w}/2)} W\left(\frac{u}{\sqrt{\pi w}}-\frac{o}{w}\right) e^{i u^2} du\right|^2.
    \label{eq:leakage_function_general}
\end{equation}
This is a powerful and general result. It shows that the spectral power of a linear chirp signal, when windowed by an arbitrary function $W$, can be expressed as the squared magnitude of a generalized Fresnel integral, where the integrand is now modulated by the window function itself, mapped into the domain of the integration variable $u$.


Finally, we extend the analysis to a real-valued signal, which is the form encountered in physical measurements. A real-valued linear chirp with unit power can be constructed from its complex analytic representation, which we denote as $s[n]$:
\begin{equation}
\begin{aligned}
    s_r[n] &= \sqrt{2}\cos\left(\phi_0 + 2\pi f_0\frac{ T}{N}n + \pi \dot{f}\frac{ T^2}{N^2} n^2\right)\\
    &= \frac{1}{\sqrt{2}}\left(s[n] + s[n]^*\right).
\end{aligned}
\end{equation}
The factor of $\sqrt{2}$ ensures that the average power of the real signal is unity. Due to the linearity of the DFT, the transform of the real signal $\tilde{s}_r[k]$, can be expressed in terms of the transform of the complex signal $\tilde{s}[k]$:
\begin{equation}
    \tilde{s}_r[k] = \frac{1}{\sqrt{2}}\left(\tilde{s}[k] + \tilde{s}[-k]^*\right).
\end{equation}
It is important to clarify the indexing notation $\tilde{s}[-k]^*$. For a DFT of length $N$, the index $-k$ is interpreted modulo $N$; that is, for $k \in \{0, \dots, N-1\}$, the index corresponds to $N-1-k$. This term represents the contribution from the complex conjugate (negative frequency) component of the signal.

The power spectrum of the real signal $\eta_r = |\tilde{s}_r[k]|^2$, is therefore:
\begin{equation}
    \eta_r(o,w) = \frac{1}{2}\left(|\tilde{s}[k]|^2 + |\tilde{s}[-k]|^2 + 2\text{Re}\{\tilde{s}[k]\tilde{s}[-k]\}\right).
\end{equation}
Substituting our leakage function $\eta(w,o) = |\tilde{s}[k]|^2$, this becomes:
\begin{equation}
    \eta_r(w,o) = \frac{1}{2}\eta(w,o) + \frac{1}{2}\eta(w, o+2f_cT) + \text{cross-term}.
\end{equation}
Here, the central time frequency of the signal is $f_c$, and the argument of the second term, $o+2f_cT$, reflects the offset from the negative-frequency center at $-f_c$. This equation has a clear physical interpretation:
\begin{enumerate}
    \item $\frac{1}{2}\eta(w,o)$: Half the power from the positive-frequency component, centered at $+f_c$.
    \item $\frac{1}{2}\eta(w,o+2f_cT)$: Half the power from the negative-frequency "mirror" component, centered at $-f_c$, which leaks into the positive frequency bins.
    \item $\text{cross-term}$: An interference term arising from the coherent sum of the positive and negative frequency components.
\end{enumerate}
When the carrier frequency is well-separated from zero frequency (i.e., $f_c \gg \dot{f}T$, or in dimensionless terms, $f_c T \gg w$), the contribution from the negative-frequency image and the cross-term become negligible for positive frequency bins ($k>0$).
\begin{equation}
    \eta_r(w,o)\approx \frac{1}{2}\eta(w,o), \quad o>0
\end{equation}
This factor of 1/2 has a direct physical interpretation. A real-valued signal is mathematically the sum of a positive-frequency complex signal and its conjugate. The total power of the unit-power real signal is equally divided between these two components. Our leakage function $\eta(w,o)$ was defined for an idealized unit-power complex signal, which has all its power concentrated at the positive frequency.

Since the positive-frequency component of the real signal contains only half the total power, its contribution to the power spectrum is correspondingly scaled by a factor of 1/2. In essence, the power measured in the positive-frequency spectrum is approximately half that of the idealized complex signal because the other half of the power resides in the negative-frequency component, which we can safely ignore in this approximation.

\section{Power-normalization of the window function}
\label{app:window_function_power}

The DFT inherently applies an implicit rectangular window to the data segment. The relationship between the time-domain power and frequency-domain power is governed by Parseval's theorem, which states:
\begin{equation}
    P = \frac{1}{N}\sum_{n=0}^{N-1}|s[n]|^2 = \sum_{k=0}^{N-1}|\tilde{s}[k]|^2.
\end{equation}
In practical STFT, however, various window functions, $W[n]$, are applied to the time series to mitigate spectral leakage. When a window is applied, the signal becomes $s[n] \cdot W[n]$, and Parseval's theorem must be applied to this new windowed sequence:
\begin{equation}
    P_W = \frac{1}{N}\sum_{n=0}^{N-1}|W[n]s[n]|^2 = \sum_{k=0}^{N-1}|\tilde{s}[k]|^2.
\end{equation}
The application of a window function inherently alters the total power of the signal. To ensure that our power measurements remain consistent and physically meaningful regardless of the window chosen, we must adopt a specific normalization convention.

For this purpose, we define a power-normalized window. Any original window function, $W_{\text{orig}}[n]$, can be converted to its power-normalized version, $W[n]$, by dividing it by its root mean square (RMS) value:
\begin{equation}
    W[n] = \frac{W_{\text{orig}}[n]}{C_{\text{power}}},
\end{equation}
where the normalization constant $C_{\text{power}}$ is the RMS value of the original window:
\begin{equation}
    C_{\text{power}} = \sqrt{\frac{1}{T}\int_{0}^{T}|W_{\text{orig}}(t)|^2 dt} \approx \sqrt{\frac{1}{N}\sum_{n=0}^{N-1}|W_{\text{orig}}[n]|^2}.
\end{equation}
By its very definition, this normalization ensures that the power of the normalized window function itself is unity $\frac{1}{N}\sum|W[n]|^2 = 1$.

The benefit of this convention becomes clear when we consider the power of a physical signal. Let's assume the amplitude of our signal $s[n]$ evolves slowly and can be approximated as constant over a single short-time Fourier transform segment. If the underlying signal has an average power of $P$, we can approximate $|s[n]|^2 \approx P$. The power of the windowed signal then becomes:
\begin{equation}
    P_W = \frac{1}{N}\sum_{n=0}^{N-1}|W[n]|^2|s[n]|^2 \approx P \left( \frac{1}{N}\sum_{n=0}^{N-1}|W[n]|^2 \right).
\end{equation}
Since our power-normalized window ensures that the term in the parenthesis is equal to one, we arrive at the desired result:
\begin{equation}
    P_W \approx P.
\end{equation}
This crucial result guarantees that the total power measured in the spectrum of the windowed signal $P_W$ remains a consistent and unbiased estimator of the true average power of the underlying physical signal $P$ regardless of the specific shape of the window function applied.

Therefore, for the gravitational wave signal discussed in our main text, we can establish a specific, window-function-independent form for the signal's power $P_i$.
\begin{equation}
    P_i \approx \frac{1}{2}h_{0,i}^2 Q_i^2.
    \label{eq:physical_power}
\end{equation}
This provides a direct physical interpretation for the power measured in the spectrum.

\section{Derivation of Critical Ratio Moments}
\label{app:weights}

In this appendix, we derive the analytical expressions for the mean ($\mu_\mathrm{CR}$) and variance ($\sigma^2_\mathrm{CR}$) of the Critical Ratio statistic, utilizing the sampling phase-averaged weighting scheme defined in \cref{eq:practical_weights}.

\subsection{Evaluation on the Real Signal Track (Perfect Match)}

To simplify the exposition, we first consider the scenario of a \textit{single-pixel track}, where for each time segment $i$, the generalized track $\mathcal{T}$ consists solely of the anchor bin (corresponding to the relative index $\kappa=0$). The generalization to multi-pixel tracks is straightforward due to the linearity of the summation.

We focus first on the numerator of the expected mean $\mu_\mathrm{CR}$, denoted as $N_\mu = \sum_{\mathcal{T}} \omega_{i,k}\lambda[i,k]$. Substituting the weight definition $\omega_{i,k} = \mathcal{L}_i \hat{\eta}_i[\kappa]$ and the factorization $\lambda[i,k] = \mathcal{L}_i \eta_i[\kappa]$, the derivation proceeds as follows:
\begin{equation}
\begin{aligned}
    N_\mu &= \sum_{i=1}^N \mathcal{L}_i \hat{\eta}_i[\kappa] \left( \mathcal{L}_i \eta_i[\kappa] \right) \\
    &= \sum_{i=1}^N \mathcal{L}_i^2 \hat{\eta}_i[\kappa] \eta_i[\kappa] \\
    &= \sum_{i=1}^N \frac{1}{2L+1} \sum_{j=i-L}^{i+L} \mathcal{L}_j^2 \hat{\eta}_j[\kappa] \eta_j[\kappa] \\
    &\approx \sum_{i=1}^N \frac{\mathcal{L}_i^2 \hat{\eta}_i[\kappa]}{2L+1} \sum_{j=i-L}^{i+L} \eta_j[\kappa] \\
    &\approx \sum_{i=1}^N \mathcal{L}_i^2 \hat{\eta}_i[\kappa] \hat{\eta}_i[\kappa] = \sum_{i=1}^N \left(\mathcal{L}_i\hat{\eta}_i[\kappa]\right)^2.
\end{aligned}
\label{eq:mu_cr_numerator_derivation}
\end{equation}
Consequently, the mean of the Critical Ratio simplifies to:
\begin{equation}
    \mu_\mathrm{CR} = \frac{1}{2}\frac{N_\mu}{\sqrt{\sum_\mathcal{T} \omega_{i,k}^2}} \approx \frac{1}{2}\sqrt{\sum_{\mathcal{T}} \left(\mathcal{L}_i\hat{\eta}_i[\kappa]\right)^2}.
    \label{eq:muCR_sim}
\end{equation}

This derivation relies on three key mathematical steps:
\begin{itemize}
    \item \textbf{Moving Average Identity (Line 2 $\to$ 3):} We rewrite the summation as a moving average over a window of $2L+1$ time segments. Neglecting boundary effects, this is a mathematical identity.
    \item \textbf{Slow-Varying Approximation (Line 3 $\to$ 4):} We assume that the macroscopic quantities---the total power statistic $\mathcal{L}_j$ and the averaged leakage factor $\hat{\eta}_j[\kappa]$---vary slowly compared to the window length $L$. They are treated as constant within the local window $[i-L, i+L]$ and factored out of the inner summation.
    \item \textbf{Ergodic Approximation (Line 4 $\to$ 5):} This is the core physical approximation. We posit that the time-average of the rapidly oscillating instantaneous leakage $\eta_j[\kappa]$ is equivalent to the phase-averaged leakage function $\hat{\eta}_i[\kappa]$ (which integrates over the offset parameter $o$). Mathematically, $\frac{1}{2L+1}\sum_{j=i-L}^{i+L} \eta_j[\kappa] \approx \hat{\eta}_i[\kappa]$. This assumes that the sequence of sampling phases $\{o_{0,j}\}$ provides a \textit{dense and uniform sampling} of the continuous offset interval $[-0.5, 0.5]$.
\end{itemize}

Following an analogous logic, the variance $\sigma^2_\mathrm{CR}$ is derived by expanding its numerator $N_\sigma = \sum_{\mathcal{T}} \omega_{i,k}^2 \lambda[i,k]$:
\begin{equation}
\begin{aligned}
    N_\sigma &= \sum_{i=1}^N \left(\mathcal{L}_i \hat{\eta}_i[\kappa]\right)^2 \left(\mathcal{L}_i \eta_i[\kappa]\right) \\
    &= \sum_{i=1}^N \frac{1}{2L+1} \sum_{j=i-L}^{i+L} \mathcal{L}_j^3 (\hat{\eta}_j[\kappa])^2 \eta_j[\kappa] \\
    &\approx \sum_{i=1}^N \frac{\mathcal{L}_i^3 (\hat{\eta}_i[\kappa])^2}{2L+1} \sum_{j=i-L}^{i+L} \eta_j[\kappa] \\
    &\approx \sum_{i=1}^N \mathcal{L}_i^3 (\hat{\eta}_i[\kappa])^2 \hat{\eta}_i[\kappa] = \sum_{i=1}^N \left(\mathcal{L}_i\hat{\eta}_i[\kappa]\right)^3.
\end{aligned}
\end{equation}
Substituting this into the general variance formula yields:
\begin{equation}
    \sigma^2_\mathrm{CR} = 1 + \frac{\sum_{\mathcal{T}} (\mathcal{L}_i\hat{\eta}_i[\kappa])^3}{\sum_{\mathcal{T}} (\mathcal{L}_i \hat{\eta}_i[\kappa])^2}.
\label{eq:sigma_cr_approx}
\end{equation}
This result highlights the dependence of variance on signal strength. In the \textit{weak-signal limit} ($\mathcal{L}_i \ll 1$), the second term vanishes, leading to $\sigma^2_\mathrm{CR} \approx 1$, consistent with the null hypothesis. For multi-pixel configurations, the derivation follows an identical logic, with the summation set $\mathcal{T}$ simply expanded to include the relevant sideband bins.

\subsection{Rationale for the 3-Pixel Track}
\label{subsec:3pixel_rationale}

To determine the optimal spatial extent of the generalized track $\mathcal{T}$, we analyze the contribution of each relative frequency bin $\kappa$ to the total detection sensitivity. Recall from \cref{eq:d_max} that the maximum detectable distance scales with the square root of the collected power term. We expand the summation over the generalized track into contributions from the central anchor bin ($\kappa=0$) and symmetric sideband pairs ($\kappa=\pm 1, \pm 2, \dots$):
\begin{widetext}
\begin{equation}
\begin{aligned}
    \sum_{(i,k) \in \mathcal{T}} \left(\mathcal{L}_i\hat{\eta}_i[\kappa]\right)^2 &= \sum_{i=1}^N \mathcal{L}_i^2 \sum_{\kappa} \hat{\eta}_i^2[\kappa] \\
    &= \sum_{i=1}^N \mathcal{L}_i^2 \left( \underbrace{\hat{\eta}_i^2[0]}_{\text{Anchor Bin}} + \underbrace{\hat{\eta}_i^2[1] + \hat{\eta}_i^2[-1]}_{\text{1st Sidebands}} + \underbrace{\hat{\eta}_i^2[2] + \hat{\eta}_i^2[-2]}_{\text{2nd Sidebands}} + \cdots \right) \\
    &\approx \sum_{i=1}^N \mathcal{L}_i^2 \underbrace{\left( \hat{\eta}^2(0, w_i) + 2\hat{\eta}^2(1, w_i)+2 \hat{\eta}^2(2, w_i)+\cdots\right)}_{\equiv \Psi(w_i)}.
\end{aligned}
\label{eq:sensitivity_expansion}
\end{equation}
\end{widetext}
Here, we invoke the \textit{slowly-varying approximation}. We assume that the expected noise power spectral density varies negligibly across the narrow frequency bandwidth (typically spanning only a few bins) where the spectral leakage function $\hat{\eta}$ exhibits significant structure. Consequently, the term $\mathcal{L}_i$ is treated as effectively constant with respect to the relative frequency index $\kappa$ and can be factored out of the summation. Proceeding with this simplification, and utilizing the symmetry of the averaged leakage function, $\hat{\eta}(o,w) = \hat{\eta}(-o,w)$, we introduce the \textit{spectral efficiency factor}, $\Psi(w)$:
\begin{equation}
    \Psi(w) \equiv \hat{\eta}^2(0,w) + 2\hat{\eta}^2(1,w)+\cdots.
    \label{eq:Psi_definition}
\end{equation}
This factor serves as a fundamental metric for the efficiency of signal power recovery in the presence of spectral leakage. It is crucial to distinguish between energy conservation and statistical sensitivity:
\begin{itemize}
    \item \textbf{Energy Conservation (Linear Sum):} The spectral leakage function itself is normalized, implying that the total signal energy is conserved regardless of how it is distributed across the frequency bins ($\sum_{\kappa} \hat{\eta}[\kappa] = 1$).
    \item \textbf{Statistical Dilution (Squared Sum):} However, the detection statistical significance scales with the \textit{sum of squares} of these factors. As the signal power disperses (smears) into more sidebands due to an increasing widening factor $w$, the sum of squares inevitably drops below unity ($\Psi(w) < 1$).
\end{itemize}
Therefore, $\Psi(w)$ represents the \textit{effective collection efficiency}. It reflects the inevitable penalty incurred by spreading signal power across multiple noise-filled frequency bins. Even if a search algorithm were to sum the signal over an infinite bandwidth to recover the power, the simultaneous accumulation of noise variance from these additional bins results in a net loss of statistical significance compared to a perfectly concentrated monochromatic signal (where $\Psi=1$).

Practical search algorithms must truncate this summation to a finite track width. The decision to restrict the analysis to the anchor bin and its immediate neighbors ($|\kappa| \le 1$, i.e., a 3-pixel track) is justified by the rapid spectral decay characteristic of standard window functions (e.g., Tukey or Hann). As illustrated in \cref{fig:eta_leakage}, the signal energy is highly localized: the central anchor bin ($\kappa=0$) captures the dominant fraction of the power, while the first sidebands ($\kappa=\pm 1$) account for the vast majority of the remaining leakage. Consequently, extending the track width beyond 3 pixels yields \textit{diminishing returns}. The contribution from higher-order sidebands ($|\kappa| \ge 2$) provides a negligible improvement in the efficiency factor $\Psi(w)$, whereas including them linearly increases the computational overhead (specifically, memory bandwidth and storage). Thus, the 3-pixel configuration represents the optimal trade-off between maximizing signal recovery fidelity and maintaining computational efficiency.

\subsection{Generalization to Template-Signal Mismatch}

The preceding analysis assumed a perfect alignment where the integer bin index $\kappa$ is defined relative to the true signal frequency. In a practical search utilizing a template bank, a continuous frequency offset inevitably exists between the template track, $f_{t,i}$, and the true signal track, $f_{r,i}$.

Let $f_{c,i}$ denote the instantaneous central frequency of the signal for the $i$-th segment. The normalized frequency offset between the true signal and the template, in units of frequency bins ($f_\mathrm{bin} = 1/T_\mathrm{DFT}$), is given by:
\begin{equation}
    o_i = \frac{f_{r,i} - f_{t,i}}{f_\mathrm{bin}}.
\end{equation}
To analyze this scenario, we distinguish between quantities derived from the \textit{template} parameters (subscript $t$) and those intrinsic to the \textit{real} signal (subscript $r$).

The detection weights, $\omega_{i,k}$, are pre-calculated based on the template trajectory:
\begin{equation}
    (\omega_{i,k})_t \propto (\mathcal{L}_i)_t (\hat{\eta}_i[\kappa])_t,
\end{equation}
where $(\hat{\eta}_i[\kappa])_t$ is the averaged leakage factor evaluated at the template's integer bin offset $\kappa$.

The actual signal contribution, $\lambda[i,k]$, however, depends on the \textit{true} signal's leakage into the bins defined by the template grid. The received power is governed by the continuous offset between the bin center and the real signal frequency. This is expressed as:
\begin{equation}
    (\lambda[i,k])_r = (\mathcal{L}_i)_r (\eta(o_i, w_i))_r.
\end{equation}
Here, we conceptualize the signal contribution in terms of the mismatch offset $o_i$. Substituting these definitions into the fundamental moments, the generalized mean of the CR becomes:
\begin{equation}
    \mu_{\mathrm{CR}} = \frac{1}{2} \frac{\sum_{\mathcal{T}} (\mathcal{L}_i \hat{\eta}_i[\kappa])_t \cdot (\mathcal{L}_i \hat{\eta}(o_i,w_i))_r}{\sqrt{\sum_{\mathcal{T}} (\mathcal{L}_i \hat{\eta}_i[\kappa])_t^2}}.
\end{equation}
Similarly, the variance generalizes to:
\begin{equation}
    \sigma^2_{\mathrm{CR}} = 1 + \frac{\sum_{\mathcal{T}} (\mathcal{L}_i \hat{\eta}_i[\kappa])_t^2 \cdot (\mathcal{L}_i \hat{\eta}(o_i,w_i))_r}{\sum_{\mathcal{T}} (\mathcal{L}_i \hat{\eta}_i[\kappa])_t^2}.
\end{equation}
These generalized expressions quantify the search performance for any arbitrary offset $o_i$. They serve as the theoretical basis for calculating the Fitting Factor (FF) and Mismatch (MM), enabling the rigorous construction of a template bank that guarantees high detection probability across the continuous parameter space.

\section{Scaling Laws and the Maximum Detectable Distance}
\label{app:dmax}
A critical characteristic of the proposed detection statistic is the scaling behavior of its moments---the mean ($\mu_\mathrm{CR}$) and variance ($\sigma^2_\mathrm{CR}$)---with respect to the number of analyzed time segments, $N$. This scaling directly governs how the search sensitivity improves with accumulating observation time.

To elucidate this behavior, we rewrite the cumulative sums over the track set $\mathcal{T}$ in terms of segment-averaged quantities. Let $\langle X \rangle_N \equiv \frac{1}{N}\sum_{i=1}^N X_i$ denote the arithmetic mean of a quantity over the $N$ segments.

First, consider the mean of the Critical Ratio. Factoring out $N$, the expression becomes:
\begin{equation}
    \mu_{\mathrm{CR}} = \frac{1}{2}\sqrt{\sum_{\mathcal{T}}\left(\mathcal{L}_i\hat{\eta}_i[\kappa]\right)^2} = \frac{1}{2}\sqrt{N \left\langle \left(\mathcal{L}_i\hat{\eta}_i[\kappa]\right)^2 \right\rangle_N}.
\end{equation}
Assuming the average signal power per segment remains roughly constant over the observation, this implies that the expected statistical significance grows with the square root of the number of segments:
\begin{equation}
    \mu_{\mathrm{CR}} \propto \sqrt{N}.
    \label{eq:mu_scaling}
\end{equation}
This $\sqrt{N}$ growth is the hallmark of semi-coherent integration strategies (often referred to as incoherent summation of coherent powers), reflecting the stochastic accumulation of signal power.

In contrast, the variance of the CR statistic exhibits a fundamentally different scaling behavior. Expressing the sums in terms of averages yields:
\begin{equation}
\begin{aligned}
    \sigma^2_{\mathrm{CR}} &= 1 + \frac{\sum_{\mathcal{T}}\left(\mathcal{L}_i\hat{\eta}_i[\kappa]\right)^3}{\sum_{\mathcal{T}}\left(\mathcal{L}_i\hat{\eta}_i[\kappa]\right)^2}\\
    &= 1 + \frac{N \left\langle \left(\mathcal{L}_i\hat{\eta}_i[\kappa]\right)^3 \right\rangle_N}{N \left\langle \left(\mathcal{L}_i\hat{\eta}_i[\kappa]\right)^2 \right\rangle_N} \\
    &= 1 + \frac{\left\langle \left(\mathcal{L}_i\hat{\eta}_i[\kappa]\right)^3 \right\rangle_N}{\left\langle \left(\mathcal{L}_i\hat{\eta}_i[\kappa]\right)^2 \right\rangle_N}.
\end{aligned}
\end{equation}
Crucially, the factor of $N$ cancels out in the fractional term. Consequently, the variance is an intensive quantity with respect to observation time. For weak signals (where the higher-order terms are small), $\sigma^2_{\mathrm{CR}}$ remains close to unity, effectively independent of the observation duration.

This divergence in scaling behaviors—the unbounded growth of the mean ($\propto \sqrt{N}$) versus the asymptotic constancy of the variance ($\propto N^0$)—is the engine of the search's sensitivity. As the observation time $N$ increases, the signal distribution shifts progressively away from the noise background without broadening significantly, thereby exponentially suppressing the false dismissal probability.

Finally, we connect this result to the astrophysical horizon. From the definition of the total power statistic, we know that $\mu_{\mathrm{CR}}$ scales quadratically with distance: $d\propto \sqrt{\mu_\mathrm{CR}}$. Combining this with \cref{eq:mu_scaling}, we arrive at the fundamental scaling law for the maximum detectable distance:
\begin{equation}
    d_{\mathrm{max}} \propto N^{1/4}.
\end{equation}
This fourth-root scaling law confirms that extending the observation duration is a robust strategy for expanding the search volume ($V \propto d^3 \propto N^{3/4}$) for mini-EMRI systems.

\section{Effective Detection Volume and Averaged Antenna Response}
\label{app:Veff}

\cref{eq:d_max} defines the maximum detectable distance $d_\mathrm{max}$ under a specific set of extrinsic parameters (sky location and polarization). However, the astrophysical performance of a search pipeline is best characterized by its sensitivity to a population of sources distributed isotropically throughout the universe. To quantify this, we transition from the horizon distance to the \textit{effective detection volume} $V_\mathrm{eff}$. This necessitates averaging the detection probability over all extrinsic variables: sky position $(\alpha, \delta)$ and source orientation $(\psi, \iota)$.

The dependence on these geometric parameters is encapsulated within the detector's antenna pattern function, $Q(t; \alpha, \delta, \psi, \iota)$. The averaging procedure begins by exploiting the separation of timescales. Since the intrinsic signal parameters (frequency and amplitude) evolve on timescales much longer than the Earth's rotation period ($T_\oplus \approx 23.93\,\mathrm{h}$), we can perform a time-average over one sidereal day. This operation effectively decouples the rapid antenna pattern modulation from the secular signal evolution. A key consequence of this averaging is that the dependence on the source's right ascension ($\alpha$) becomes degenerate, reducing the effective sky-dependence to a function of declination ($\delta$) only.

Formally, the effective volume is constructed by integrating the detectable volume element over the full parameter space. We first define the volume for a fixed source orientation, $V(\psi, \iota)$, by integrating over the celestial sphere:
\begin{equation}
\begin{aligned}
    V(\psi,\iota) &= \int_0^{2\pi} d\alpha \int_{-1}^{1} d(\cos\delta) \int_{0}^{d_\mathrm{max}(\delta, \psi,\iota)} r^2 dr \\
    &= \frac{1}{3} \int_0^{2\pi} d\alpha \int_{-1}^{1} d(\cos\delta) \, d_\mathrm{max}^3(\delta, \psi,\iota).
\end{aligned}
\end{equation}
The final effective detection volume, $V_\mathrm{eff}$, is obtained by averaging this quantity over all possible polarization angles $\psi$ and inclination angles $\iota$:
\begin{equation}
    V_\mathrm{eff} = \frac{1}{4\pi}\int_0^{2\pi} d\psi \int_{-1}^{1} d(\cos\iota) \, V(\psi,\iota).
\end{equation}

\begin{widetext}
Combining these integrals yields the complete expression for the effective detection volume:
\begin{equation}
    V_\mathrm{eff} = \frac{1}{12\pi} \int_0^{2\pi} d\psi \int_{-1}^{1} d(\cos\iota)\int_0^{2\pi} d\alpha \int_{-1}^{1} d(\cos\delta) \, d_\mathrm{max}^3(\delta, \psi,\iota).
    \label{eq:Veff_full_integral}
\end{equation}
This scalar metric provides a robust quantification of the pipeline's global sensitivity. It is often convenient to express this volume in terms of an \textit{effective distance}, $d_\mathrm{eff}$, defined as the radius of a Euclidean sphere with volume $V_\mathrm{eff}$:
\begin{equation}
    d_\mathrm{eff} = \left(\frac{3 V_\mathrm{eff}}{4\pi}\right)^{1/3}.
    \label{eq:deff_definition}
\end{equation}

While \cref{eq:Veff_full_integral} is exact, it is computationally intensive. We can simplify it conceptually and practically by encapsulating the entire geometric averaging process into a single, scalar factor: the \textit{fully-averaged antenna pattern factor}, denoted as $\langle Q \rangle$. This factor represents the detector's effective response to an isotropic source population. For the $\Sigma R$ statistic (where sensitivity scales as the fourth root of power accumulation, see \cref{app:dmax}), this factor is defined as:
\begin{equation}
    \langle Q \rangle_{\Sigma R} = \left\langle \langle Q(t)^4 \rangle_t^{3/4} \right\rangle_{\mathrm{sky,orient}}^{1/3}.
    \label{eq:Q_averaged_SigmaR}
\end{equation}
Numerical evaluation of this integral yields the following characteristic values for the current detector network:
\begin{equation}
\begin{aligned}
    \langle Q \rangle_{\Sigma R}^{\text{LIGO H}} &\approx 0.4590, \quad
    \langle Q \rangle_{\Sigma R}^{\text{LIGO L}} \approx 0.4604, \quad
    \langle Q \rangle_{\Sigma R}^{\text{Virgo}} \approx 0.4597.
    \label{eq:Q_averaged}
\end{aligned}
\end{equation}

By factoring out this geometric term, the effective distance can be expressed in a modular form that separates geometry from signal strength:
\begin{equation}
    d_\mathrm{eff} = \langle Q \rangle_{\Sigma R} \left[ \sqrt{2}~\mathrm{erfc}^{-1}(2P_{\mathrm{fa}}) + \sqrt{2}~\mathrm{erfc}^{-1}(2P_{\mathrm{fd}}) \right]^{-1/2}  \left[ \frac{1}{4}\sum_{\mathcal{T}} h_{e,i}^4 \frac{T_\mathrm{DFT}^2}{S_{n,i}^2}\hat{\eta}_i^2[\kappa]\right]^{1/4}.
    \label{eq:deff_compact}
\end{equation}

For comparison, the conventional Hough transform pipeline exhibits a different sensitivity scaling due to its binary thresholding nature. The corresponding averaged factor for the Hough transform is derived as:
\begin{equation}
    \langle Q \rangle_{\mathrm{Hough}} = \left\langle \langle Q(t)^2 \rangle_t^{3/2} \right\rangle_{\mathrm{sky,orient}}^{1/3}.
    \label{eq:Q_averaged_Hough}
\end{equation}
Numerical integration yields effective response factors for the Hough pipeline: $\langle Q \rangle_{\mathrm{Hough}}^{\text{LIGO H}} \approx 0.4207$, $\langle Q \rangle_{\mathrm{Hough}}^{\text{LIGO L}} \approx 0.4191$, and $\langle Q \rangle_{\mathrm{Hough}}^{\text{Virgo}} \approx 0.4198$.
\end{widetext}

\section{Frequency layering strategy}
\label{app:Freq_layer}

To rigorously determine the optimal frequency layering configuration, we first establish a continuous metric for the accumulation of statistical significance. We observe that the squared expectation of the Critical Ratio, $\mu_\mathrm{CR}^2$, behaves as an additive quantity. For a search trajectory $\mathcal{T}$ composed of disjoint sub-tracks $\mathcal{T}_1$ and $\mathcal{T}_2$, the total significance satisfies:
\begin{equation}
    \mu_\mathrm{CR}^2(\mathcal{T}_1+\mathcal{T}_2) = \mu_\mathrm{CR}^2(\mathcal{T}_1) + \mu_\mathrm{CR}^2(\mathcal{T}_2).
\end{equation}
Substituting the definition of $\mu_\mathrm{CR}$ from \cref{eq:CR_moments_practical} and expanding the terms for the 3-pixel track configuration, the total squared significance over $N$ segments is:
\begin{equation}
\begin{aligned}
    \mu_\mathrm{CR}^2 &= \frac{1}{4}\sum_{(i,k)\in\mathcal{T}}\left(\mathcal{L}_i\hat{\eta}_i[\kappa]\right)^2 \\
    &= \frac{1}{4}\sum_{i=1}^N \mathcal{L}_i^2 \left(\hat{\eta}_i^2[0] + 2\hat{\eta}_i^2[1]\right) \\
    &= \frac{1}{4}\sum_{i=1}^N h_{0,i}^4 Q_i^4\frac{  T_\mathrm{DFT}^2}{S_{n,i}^2} \left[\hat{\eta}^2(0,w_i) + 2\hat{\eta}^2(1,w_i)\right].
\end{aligned}
\end{equation}
Here, we assume the standard condition of 50\% temporal overlap between adjacent segments. In the \textbf{continuum limit}, the summation over discrete segments can be approximated by a time integral, where the segment density is given by $dN/dt = (T_\mathrm{DFT}/2)^{-1}$. Applying the fully-averaged antenna pattern factor $\langle Q \rangle_{\Sigma R}$, the expression transforms to:
\begin{equation}
\begin{aligned}
    \mu_\mathrm{CR}^2 &\approx \frac{1}{4} \langle Q \rangle_{\Sigma R}^4 \int \frac{h_{0}^4(t) T_\mathrm{DFT}^2}{S_{n}^2(f(t))} \Psi(w(t)) \frac{2}{T_\mathrm{DFT}} dt \\
    &= 2 \langle Q \rangle_{\Sigma R}^4 \int h_{0}^4(f(t)) \frac{T_\mathrm{DFT}}{S_{n}^2(f(t))} \Psi(w(t)) \, dt,
\end{aligned}
\end{equation}
where $\Psi(w) \equiv \hat{\eta}^2(0,w) + 2\hat{\eta}^2(1,w)$ represents the \textit{spectral efficiency factor} for the default 3-pixel track configuration, as formally defined in \cref{eq:Psi_definition}.

Finally, to facilitate optimization in the frequency domain, we perform a change of variables from time $t$ to frequency $f$. Using the stationary phase relationship $dt = (df/dt)^{-1} df = \dot{f}^{-1} df$, we arrive at the spectral representation of the sensitivity:
\begin{equation}
    \mu_\mathrm{CR}^2 = \frac{1}{4} \langle Q \rangle_{\Sigma R}^4 \int_{f_\mathrm{min}}^{f_\mathrm{max}} h_{0}^4(f)\frac{T_\mathrm{DFT}}{S_{n}^2(f)} \Psi\left(\dot{f}T_\mathrm{DFT}^2\right) \dot{f}^{-1}\, df.
\end{equation}

Based on this integral form, we can formally define the \textit{Sensitivity Density} $\rho_\mathrm{CR}(f, T_\mathrm{DFT})$, which quantifies the contribution of each unit frequency interval to the total squared detection statistic:
\begin{equation}
\begin{aligned}
    \rho_\mathrm{CR}(f, T_\mathrm{DFT}) &\equiv \frac{d(\mu_\mathrm{CR}^2)}{df} \\
    &=\frac{1}{4} \langle Q \rangle_{\Sigma R}^4\,h_{0}^4(f)\frac{T_\mathrm{DFT}}{S_{n}^2(f)} \Psi\left(\dot{f}T_\mathrm{DFT}^2\right) \dot{f}^{-1}.
\end{aligned}
    \label{eq:CR_density}
\end{equation}
This density function reveals the fundamental trade-off: increasing $T_\mathrm{DFT}$ linearly increases the SNR accumulation (the linear $T_\mathrm{DFT}$ term), but eventually suppresses the efficiency factor $\Psi(w)$ as the widening $w = \dot{f}T_\mathrm{DFT}^2$ grows large. The optimal frequency layering strategy is thus equivalent to finding the function $T_\mathrm{DFT}(f)$ that maximizes this density at every frequency $f$.

For a monotonically evolving signal (where the frequency $f(t)$ is a strictly increasing function of time), the total time-frequency trajectory can be uniquely partitioned into $N$ disjoint frequency layers. Consider a search configuration partitioned into $N$ distinct frequency layers. This segmentation is defined by a set of $N-1$ critical frequencies, denoted as $\{f_{c,1}, f_{c,2}, \dots, f_{c,N-1}\}$, which divide the total bandwidth $[f_\mathrm{min}, f_\mathrm{max}]$ into contiguous intervals. Within the $n$-th layer, a specific coherent integration time $T_n$ is applied.

The total squared statistical significance is obtained by summing the integral contributions from each layer:
\begin{equation}
\begin{aligned}
    \mu_{\mathrm{CR}}^2 &= \sum_{n=1}^{N} \mu_{\mathrm{CR}}^2(\text{layer } n) \\
    &= \int_{f_\mathrm{min}}^{f_{c,1}} \rho_{\mathrm{CR}}(f, T_1) \, df + \int_{f_{c,1}}^{f_{c,2}} \rho_{\mathrm{CR}}(f, T_2) \, df + \cdots \\
    &\quad+ \int_{f_{c,N-1}}^{f_\mathrm{max}} \rho_{\mathrm{CR}}(f, T_{N}) \, df.
\end{aligned}
\label{eq:mu_CR_total_layered}
\end{equation}
This formulation transforms the optimization problem into finding the optimal set of transition frequencies $\{f_{c,n}\}$ and coherent times $\{T_n\}$ that maximize this global sum. 

It is straightforward to show that the optimal critical frequency, $f_{c,n}$, separating two adjacent layers (with integration times $T_n$ and $T_{n+1}$) is determined \textit{solely} by the signal's frequency derivative $\dot{f}$.

The optimal transition point occurs where the sensitivity densities of the two configurations intersect, i.e., where switching from $T_n$ to $T_{n+1}$ yields no discontinuities in sensitivity:
\begin{equation}
    \rho_{\mathrm{CR}}(f_{c,n}, T_n) = \rho_{\mathrm{CR}}(f_{c,n}, T_{n+1}).
\end{equation}
Substituting the definition of $\rho_{\mathrm{CR}}$ from \cref{eq:CR_density}, we observe that the terms depending on the signal strength $h_0(f)$ and the detector noise $S_n(f)$ appear as common factors on both sides and explicitly cancel out. The condition simplifies to a transcendental equation involving only the integration times and the frequency derivative:
\begin{equation}
    T_n \Psi\left(\dot{f}(f_{c,n}) T_n^2\right) = T_{n+1} \Psi\left(\dot{f}(f_{c,n}) T_{n+1}^2\right).
    \label{eq:critical_freq_condition}
\end{equation}
This result reveals a fundamental universality in the search design: the optimal frequency layering is governed exclusively by the interplay between the coherent gain (linear in $T$) and the spectral leakage (encapsulated in $\Psi$), independent of the specific spectral shape of the noise or the source.

\section{Statistical Independence and Pixel Correlations}
\label{app:correlations}

In the derivation of the $\Sigma R$ statistic (\cref{eq:sum_R_mean_variance}), we posited the simplifying assumption that the power ratios $R[i,k]$ behave as mutually independent random variables across the entire time-frequency plane. While this hypothesis affords analytical tractability, it represents an idealization. In practice, the inherent mathematical structure of the STFT inevitably induces non-vanishing statistical correlations between pixels $(i,k)$ and $(j,l)$ with sufficiently small separation. These correlations arise from two primary mechanisms:
\begin{enumerate}
    \item \textbf{Spectral Correlations (Frequency Axis):} The application of a window function $W[n]$ imposes an amplitude modulation on the time-domain data. In the frequency domain, this manifests as spectral leakage: the windowing process convolves the noise spectrum with the window's kernel, thereby smearing noise power across adjacent bins.
    \item \textbf{Temporal Correlations (Time Axis):} These correlations arise primarily from the use of overlapping time segments (e.g., 50\% overlap), where adjacent STFT columns share common time-domain data samples. Furthermore, even in the absence of physical overlap, the intrinsic temporal structure of the detector noise (i.e., its non-white nature and finite correlation length) introduces residual statistical coupling between consecutive time segments.
\end{enumerate}

To rigorously characterize these effects, we construct the covariance structure of the spectrogram. For stationary Gaussian noise, the correlation in the time-frequency plane is mathematically governed by the properties of the window function and the segment overlap.

First, we examine the correlation between the complex DFT spectrum of the $i$-th and $j$-th time segments, denoted as $\tilde{x}_i[k]$ and $\tilde{x}_j[l]$. The expectation of their cross-product is:
\begin{widetext}
\begin{equation}
\begin{aligned}
    \mathbb{E}\big[\tilde{x}_i[k] \tilde{x}_j^*[l]\big] &= \mathbb{E}\left[ \left(\frac{1}{N}\sum_{n=0}^{N-1} W[n] x_i[n] e^{-i \frac{2 \pi}{N} k n}\right) \left(\frac{1}{N}\sum_{m=0}^{N-1} W[m] x_j[m] e^{i \frac{2 \pi}{N} l m}\right)^* \right] \\
    &= \frac{1}{N^2}\sum_{n=0}^{N-1} \sum_{m=0}^{N-1} W[n] W[m] \mathbb{E}\big[x_i[n] x_j[m]\big] e^{-i \frac{2 \pi}{N} (k n - l m)}.
\end{aligned}
\end{equation}
\end{widetext}
Let $R_{ij}[n,m] \equiv \mathbb{E}[x_i[n] x_j[m]]$ denote the cross-correlation matrix of the time-domain noise between segments $i$ and $j$. We further define a windowing mask matrix, $M[n,m] \equiv W[n]W[m]$.
We can define two auxiliary spectral correlation matrices: $C_{ij}$ (conjugate covariance) and $D_{ij}$ (non-conjugate covariance/relation):
\begin{equation}
\begin{aligned}
    C_{ij}[k,l] &\equiv \mathbb{E}\big[\tilde{x}_i[k] \tilde{x}_j^*[l]\big], \\
    D_{ij}[k,l] &\equiv \mathbb{E}\big[\tilde{x}_i[k] \tilde{x}_j[l]\big].
\end{aligned}
\end{equation}
These terms can be elegantly expressed as spectrum of the 2D DFT of the windowed time-domain correlation matrix:
\begin{equation}
\begin{aligned}
    C_{ij}[k,l] &= \mathrm{DFT2}\left\{ M \odot R_{ij} \right\}[k, -l], \\
    D_{ij}[k,l] &= \mathrm{DFT2}\left\{ M \odot R_{ij} \right\}[k, l],
\end{aligned}
\end{equation}
where $\odot$ denotes the element-wise (Hadamard) product:
\begin{equation}
    (M \odot R_{ij})[n,m] = M[n,m]\cdot R_{ij}[n,m].
\end{equation}

We now derive the covariance of the power ratio $R$ across distinct time-frequency bins. We first consider the case of pure noise. Our goal is to calculate the covariance $\mathrm{Cov}(R[i,k], R[j,l])$.

By definition, the covariance of the normalized variables is scaled by their respective normalization factors:
\begin{equation}
    \mathrm{Cov}\left(R[i,k], R[j,l]\right) = \frac{ \mathrm{Cov}\left(|\tilde{n}_i[k]|^2, |\tilde{n}_j[l]|^2\right)}{\langle|\tilde{n}_i[k]|^2\rangle \langle|\tilde{n}_j[l]|^2\rangle}.
\end{equation}
To evaluate the numerator, we recall the definition $\mathrm{Cov}(X,Y) = \mathbb{E}[XY] - \mathbb{E}[X]\mathbb{E}[Y]$. We invoke \textit{Wick's Theorem} for zero-mean complex Gaussian variables to decompose the fourth-order moment $\mathbb{E}[|\tilde{n}_i|^2 |\tilde{n}_j|^2]$ into the sum of pairwise expectations:
\begin{equation}
\begin{aligned}
    \mathbb{E}\left[|\tilde{n}_i|^2 |\tilde{n}_j|^2\right] =&\  \mathbb{E}[|\tilde{n}_i|^2]\mathbb{E}[|\tilde{n}_j|^2] \\
    &+ \left|\mathbb{E}(\tilde{n}_i \tilde{n}_j^*)\right|^2 + \left|\mathbb{E}(\tilde{n}_i \tilde{n}_j)\right|^2.
\end{aligned}
\end{equation}
When substituting this expansion back into the covariance expression, the first term on the right-hand side ($\mathbb{E}[|\tilde{n}_i|^2]\mathbb{E}[|\tilde{n}_j|^2]$) exactly cancels with the product of the means. This leaves only the cross-correlation terms. Using the $C_{ij}$ and $D_{ij}$ notation defined previously, we arrive at the final generalized covariance expression:
\begin{equation}
    \mathrm{Cov}\left(R[i,k], R[j,l]\right) = \frac{\left| C_{ij}[k,l] \right|^2 + \left| D_{ij}[k,l] \right|^2}{\langle|\tilde{n}_i[k]|^2\rangle \langle|\tilde{n}_j[l]|^2\rangle}.
    \label{eq:R_covariance_matrix}
\end{equation}
The above derivation assumes pure noise. In the presence of a signal ($\tilde{s} \neq 0$), $R$ follows a \textit{non-central} distribution. While exact analytical expressions for the covariance in the signal-present regime can be derived following an analogous methodological framework—specifically by expanding the fourth-order moments of the non-central variables—the resulting formulae are algebraically prohibitive to display. They involve multiple cross-terms representing the interaction between the deterministic signal amplitude and the stochastic noise background. Consequently, we omit these explicit forms for the sake of brevity. However, fundamentally, the correlation structure remains governed by the same windowing and overlap mechanisms identified in the noise-only analysis.

Finally, we explicitly quantify the impact of these correlations on the statistics of the aggregate metric $\Sigma R$. While the linearity of the expectation operator ensures that the mean $\mu_{\Sigma R}$ remains unbiased regardless of correlations, the variance $\sigma^2_{\Sigma R}$ requires a correction term to account for the covariance between contributing pixels:
\begin{widetext}
\begin{equation}
\begin{aligned}
    \sigma^2_{\Sigma R} &= \sum_{(i,k)\in\mathcal{T}} \omega_{i,k}^2\sigma^2_{R}[i,k] + \sum_{\text{distinct pairs}} 2\omega_{i,k}\omega_{j,l}\mathrm{Cov}(R[i,k],R[j,l]) \\
    &= \sigma^2_{\Sigma R,0} + \Delta\sigma^2_{\mathrm{corr}}.
\end{aligned}
\label{eq:variance_correction}
\end{equation}
\end{widetext}
Here, the second summation runs over all distinct pairs of pixels in the track set. The factor of 2 explicitly accounts for the symmetry of the covariance matrix ($\mathrm{Cov}(X,Y) = \mathrm{Cov}(Y,X)$). The term $\sigma^2_{\Sigma R,0}$ denotes the variance derived under the independence assumption, while the second term captures the cumulative contribution of cross-correlations. Since the windowing-induced correlations are predominantly positive, this correction term leads to \textit{variance inflation} (i.e., $\sigma^2_{\Sigma R} > \sigma^2_{\Sigma R,0}$).

In the regime of weak correlations (e.g., standard 50\% temporal overlap), this inflation is minor. Consequently, for the calculation of the maximum detectable distance $d_{\mathrm{max}}$, where we adopt the approximation $\sigma_{\mathrm{CR}} \approx 1$, the impact of this correction is second-order and can be safely neglected. However, \cref{eq:variance_correction} serves as a crucial cautionary principle: it dictates that one cannot arbitrarily enhance sensitivity by simply increasing the density of STFT segments within a fixed observation duration. Excessive temporal overlap (e.g., $\sim 100\%$) would drastically increase the number of correlated cross-terms. This would cause the background variance to grow faster than the signal accumulation, leading to diminishing returns or even a degradation in the statistical significance.

\bibliography{apssamp}

\end{document}